\documentclass[12pt]{article}

\makeatletter
\newcommand{\stars}{* $p<0.1$, ** $p<0.05$, *** $p<0.01$. }

\newcommand{\stderr}{Robust standard errors are in parentheses. }

\newcommand{\sampleperiod}{January 1970 to December 2019 }
\def\equationautorefname~#1\null{Equation (#1)\null}
\usepackage{amssymb,amsmath,amsfonts,eurosym,geometry,ulem,graphicx,caption,color,setspace,sectsty,comment,footmisc,caption,pdflscape,array}
\usepackage{bbm}
\usepackage{pdflscape} 
\usepackage{afterpage} 
\usepackage{float}
\usepackage{tikz}
\usepackage{pgffor}
\usepackage{caption}
\usepackage{neuralnetwork}
\usepackage{booktabs}
\usepackage{pdflscape}
\usepackage{xstring}
\usepackage{authblk}
\usepackage{multirow}
\usepackage{subcaption}
\newcolumntype{L}[1]{>{\raggedright\let\newline\\arraybackslash\hspace{0pt}}m{#1}}
\newcolumntype{C}[1]{>{\centering\let\newline\\arraybackslash\hspace{0pt}}m{#1}}
\newcolumntype{R}[1]{>{\raggedleft\let\newline\\arraybackslash\hspace{0pt}}m{#1}}

\usepackage{color}
\usepackage{amsthm}
\usepackage[authoryear]{natbib}
\usepackage{setspace}
\usepackage[unicode=true,
 bookmarks=true,bookmarksnumbered=false,bookmarksopen=false,
 breaklinks=false,pdfborder={0 0 1},backref=false,colorlinks=true]
 {hyperref}

\usepackage[textsize=tiny]{todonotes}

\usepackage{longtable}
\usepackage{siunitx}
\usepackage[flushleft]{threeparttable}

\makeatother

\usetikzlibrary{shapes,arrows,positioning,chains}

\usepackage{array} 

\DeclareRobustCommand{\rvdots}{%
  \vbox{
    \baselineskip4\p@\lineskiplimit\z@
    \kern-\p@
    \hbox{.}\hbox{.}\hbox{.}
  }}
  
\makeatother

\hypersetup{pdftitle={Deep Learning for Conditional Asset Pricing Models},
	pdfauthor={Hongyi Liu},
	colorlinks=true,linkcolor=blue!50!black,citecolor=blue!50!black}

\begin{document}

\begin{titlepage}
\title{\vspace{-3.5cm}Deep Learning for Conditional Asset Pricing Models \thanks{We are grateful for valuable advice from Andreas Neuhierl and Werner Ploberger on this paper. We would like to thank Caio Ibsen Rodrigues de Almeida, Zhikun Cai (discussant), Qihui Chen, Jerome Detemple, Andrew Ellul, Junyi Gao (discussant), Christian Heyerdahl-Larsen, Xing Huang, Raymond Leung (discussant), Yingying Li, Garcia Ren\'{e}, Jay Ritter, Andrea Vedolin, Liqun Yu (discussant), Jie Xu, and Fernando Zapatero for helpful comments. This paper also benefited from the comments of seminar audiences  or conference participants at Boston University, HKUST (Guangzhou), and Washington University in St. Louis, and 2021 Asian \& North American Summer Meeting of the Econometric Society. Discussions with Robert Korajczyk on related projects greatly enlightened me about this project. Any findings, and conclusions or recommendations expressed in this paper are those of the authors.}}
\author{Hongyi Liu\thanks{Washington University in St. Louis, Department of Economics, Email: hongyi.liu@wustl.edu} }
\date{June, 2021}


\maketitle

\begin{abstract}
We propose a new pseudo-Siamese Network for Asset Pricing (SNAP) model, based on deep learning approaches, for conditional asset pricing. Our model allows for the deep alpha, deep beta and deep factor risk premia conditional on high dimensional observable information of financial characteristics and macroeconomic states, while storing the long-term dependency of the informative features through long short-term memory network. We apply this method to monthly U.S. stock returns from 1970-2019 and find that our pseudo-SNAP model outperforms the benchmark approaches in terms of out-of-sample prediction and out-of-sample Sharpe ratio. In addition, we also apply our method to calculate deep mispricing errors which we use to construct an arbitrage portfolio K-Means clustering. We find that the arbitrage portfolio has significant alphas.
\vspace{0in}


\bigskip
\end{abstract}
\setcounter{page}{0}
\thispagestyle{empty}
\end{titlepage}
\pagebreak \newpage

\begin{singlespace}
	\begin{quote}
		``\textit{In the short run, the market is a voting machine, but in the long run it is a weighing machine}.''
		\hfill  -- Benjamin Graham 
	\end{quote}
\end{singlespace} 
\doublespacing

\section{Introduction} \label{sec:introduction}
A fundamental task in the asset pricing literature is to explain why average returns vary across assets and whether different returns are priced by different risk exposures of the assets (beta) and asset mispricing (alpha). However, the prediction of stock excess returns is one of the challenging questions for financial economist and practitioners. \cite{simin_poor_2008} shows the poor predictive performance of asset pricing models by analyzing the forecast errors from the CAPM \citep{sharpe_capital_1964,lintner_valuation_1965} and Fama-French 3-factor model \citep{fama_common_1993}. This distinction leads the trade-off between interpretation and prediction of asset pricing models. In fact, although hundreds of research papers have been dedicated to enhance interpretability of asset pricing models \citep{harvey__2016}, the empirical asset pricing literature lacks a discussion of how to improve model predictability.

To solve this problem, we implement a pseudo-Siamese neural network to connect the sub-networks of beta, alpha, and factor risk premia in a network. The pseudo-Siamese network here allows us to estimate beta, alpha, and factor risk in a one-pass framework, which classifies the effects among these three pricing elements to asset returns in an adversarial manner. More importantly, such an adversarial manner allows us to refine the out-of-sample predictive ability at each training step by evaluating the mean square errors between one-month forward excess returns and its predictor estimated from our pseudo-SNAP model. In addition, different from the autoencoder asset pricing models by \cite{gu_autoencoder_2020}, the long short-term memory network in our pseudo-Siamese architecture, is capable of storing the long-run memory about the sequence of firm characteristics and macroeconomic variables along with the most recent network (short-run memory) to further improve the out-of-sample predictive performance of our model. In practice, Benjamin Graham summarized the stock market with the above quote, which expresses the distinct stance between short-run and long-run stock market behavior. However, few of research has ever tested or verified this statement. We construct an arbitrage portfolio based on the estimated alphas from our model and verify this quote in accordance with the Sharpe ratio of this portfolio. This exercise aims to respond to the question of whether our prediction-improved model is able to explain to some extent the long-run behavior of investors in practice.

Why do traditional asset pricing models predict poorly? One cause is due to the strong assumptions on asset pricing models. There are typically two approaches in identifying return predictors. One leading approach is from the seminal contribution of \cite{fama_common_1993, fama_five-factor_2015}, which framed factor-mimicking portfolios based on sorting multiple firm characteristics to predict stock returns.  Likewise, much of the recent research in empirical asset pricing uses characteristics-based factors to capture average returns while treating the intercept as the mispricing alpha \citep{carhart_persistence_1997, hou_digesting_2015, feng_taming_2020}. Another approach is using linear regression as applied by \cite{fama_risk_1973}. However, betas and alphas in these studies are assumed to be time-invariant and independent of firm characteristics. As many researchers have argued the most explicit failure is within empirical asset pricing, such as CAPM, which can be attributed to the absence of conditional information of the model in previous studies \citep{harvey_time-varying_1989, ferson_variation_1991,jagannathan_conditional_1996,ferson_conditioning_1999,lettau_resurrecting_2001,nagel_estimation_2011}. Furthermore, the sorting portfolio is fragile due to the curse of dimensionality (when the dimension of firm characteristics is larger than the number of observations)\footnote{When the number of firm characteristics grows quickly for a given number of observations, the data is unable to keep up and becomes sparse.}. Another issue is that linear regression often requires strong assumptions on functional forms which may lead to the loss of some economic interpretations, such as non-linearity and data interactions. 

Another cause is that the effect of existing factors will be attenuated or disappeared after the factors made their debut. \cite{schwert_chapter_2003} points out that the effects of size and value factors are decayed or gone. Furthermore, recently \cite{hou_replicating_2020} replicate 452 anomaly variables and find that most of them fail to replicate. Such a performance decay issue remains an open question.

To resolve the above questions, we apply long short-term memory networks (LSTM, hereafter) \citep{hochreiter_long_1997} particularly, a special type of recurrent neural networks (RNN) in deep learning, to formulate the time-varying risk factors $ \lambda_{t} $, betas $ \beta_{it}$, and alphas $ \alpha_{it} $ while remaining agnostic to functional form. Sequential data, such as financial time series data, often encompasses a mixture of long-term and short-term dependencies, while traditional time series approaches, such as autoregressive models, fail to capture this pattern. LSTM has the ability to learn intricate long-term temporal dependencies in sequential data and has achieved state-of-art performance on many difficult tasks, including handwriting recognition, language modeling, acoustic modeling of speech, and macroeconomic indicator forecasting \citep{greff_lstm_2017, smalter_hall_macroeconomic_2017}. That is to say, financial time series data has memories after being formulated by LSTM. Due to the deep network representation, $ \lambda_{t} $, $ \beta_{it}$, and $ \alpha_{it} $ are named after deep risk factors, deep betas, and deep alphas in our paper. To characterize key signaling variables for predicting stock excess returns, we conduct the task of extracting meaningful financial and macroeconomic pricing information from high dimensional informative data spaces. In particular, we consider 90 firm characteristics, $ c_{it} $ as input for characteristic-based $ \beta_{it}$ and $ \alpha_{it} $ LSTM networks, respectively, and a total of 215 variables (including 90 average firm characteristics, 124 macroeconomic variables, and market excess returns) as input for risk factors. The deep neural network allows us to produce effective training models from large architectures. This property is confirmed by the universal representation theorem which states that any shallow feedforward neural network with large enough hidden units can approximate all continuous functions \citep{hornik_multilayer_1989}.  In other words, deep learning methods are capable of fitting complex functions with high dimensional variables which are difficult to be specified from an economic model.


In a study related to our paper, \cite{kelly_characteristics_2019} propose an instrumented principal component analysis (IPCA) approach to allow  $ \beta_{it}$ and $ \alpha_{it} $  to be instrumented by firm characteristics. Similarly, \cite{kim_arbitrage_2020} allow both $\beta_{it}$ and $ \alpha_{it} $ to be functions of firm characteristics and estimate them using projected-principal component analysis (PPCA). However, these studies impose restrictions on representing $\beta_{it}$ and $ \alpha_{it} $ by the parametric linear additive functions of firm characteristics. Also, principal component analysis (PCA) does a poor job in out-of-sample forecasting \citep{hastie_elements_2009,goodfellow_deep_2016}. In addition, as an extension to \cite{connor_efficient_2012}, \cite{li_dynamic_2020} assume $ \beta_{it}$ and $ \alpha_{it} $ are additive semi-parametric functions of firm characteristics. However, \cite{schmidt-hieber_nonparametric_2020} show that multivariate nonparametric regression models estimated by the deep neural network with ReLU activation function achieve the minimax rates of convergence and \cite{poggio_deep_2016} points out that deep neural networks have a superior performance compared to linear additive models. Another related predecessor mentioned earlier, \cite{gu_autoencoder_2020} implement the autoencoder neural network to estimate nonlinear conditional risk exposures along with firm characteristics and to estimate the risk factors along with returns. This model, however, fails to capture mispricing and only allows either risk exposures to factor or risk factors at time $ t $ to be encoded by the contemporaneous information set. Moreover, an autoencoder, is mainly used for feature extraction or dimensionality reduction, but it could lose important information in the input features. Because an autoencoder tends to learn as much information as possible rather than as much pertinent information as possible.

Notice that our paper aims to combine deep learning methodologies with the conditional asset pricing model in an economic theory-based configuration rather than feeding all of the informative variables, such as firm characteristics or macroeconomic variables, in a ``black box" to predict expected returns. In understanding the economic principle it is important to comprehend pricing models, because it helps against model fishing. To do so, we first derive the conditional single-beta representation of one-factor model with a general form of a nonlinear stochastic discount factor (SDF). Then we formulate the deep alphas, deep betas, and the deep risk factors as the LSTM network function of the conditional informational set(s) (firm characteristics and macroeconomic states mentioned earlier). Finally, we concatenate the stock excess returns with these LSTM sub-networks via pseudo-Siamese network. Hence our pseudo-Siamese network with LSTM is set up under the conditional asset pricing models as derived in section \ref{sec:model}. 

To evaluate whether the deep mispricing errors (deep alphas) are zero, we propose a novel way of testing zero mispricing errors. In particular, we simplify the mispricing hypothesis testing into a controlled experiment testing question. In simple words, we take the pseudo-SNAP model as a control group, while the experimental group, referred to as the masked pseudo-SNAP model, is taken by masking out the deep alphas sub-network only with the other sub-networks held constant. The hypothesis testing of $ \alpha_{it} = 0$ is equivalent to the group difference testing question between the residuals of the pseudo-SNAP model and the masked alpha network. Then we implement the Mann-Whitney U test statistic to test it. The mispricing testing significantly rejects the null hypothesis that the mispricing errors are zero.
Following the setting in \cite{kim_arbitrage_2020}, we construct alpha-weighted arbitrage portfolios and find the significant alphas across three classical asset pricing models (CAPM, Fama-French 3-factor and 5-factor model). We further use K-means, an unsupervised learning method, to classify five types of  arbitrage portfolios in each month from January 2006 to December 2019. Our K-Means arbitrage portfolios show that there are always popular and unpopular portfolios with high positive or high negative Sharpe ratio over time, but the Sharpe ratio difference between the distinct types of portfolios gradually decreases over the long term, which provides an empirical verification for Benjamin Graham's quote.

Our paper analyzes 50 years of individual stock returns in the U.S. and finds that our pseudo-Siamese Network for Asset Pricing (SNAP) model outperforms the benchmark models including: Fama-French 3 and 5 factor model, and LASSO, ElasticNet, ridge regression, and feedforward network. The criterion of model comparison is based on the predictive $ R^2 $ and the out-of-sample Sharpe ratio $ SR $. The former describes the fraction of variation of stock returns one-month later. This variation is explained by current realized return predictors, and the $ SR $ measures the performance of a portfolio compared to a risk-free asset. Our contribution is five-fold. First, we propose a pseudo-SNAP model for conditional asset pricing. Our model builds up the bridge between the deep learning literature and conditional asset pricing models, and our pseudo-SNAP model outperforms the benchmark models in terms of out-of-sample prediction and out-of-sample Sharpe ratio. In particular, our model yields a stellar out-of-sample Sharpe ratio above 2 and achieves the smallest performance decay compared to other benchmarks. The performance is robust under economic restrictions. Empirically, we also find that the top 20 influential characteristics and macroeconomic variables are important for predicting stock returns. Second, we also underscore the importance of incorporating the deep learning approaches to conditional asset pricing models. In other words, a correctly specified asset pricing model with deep learning approaches is the key to achieving state-of-the-art performance. Third, we propose a novel way of testing mispricing errors through simplifying the hypothesis $ \alpha_{it} = 0$ into a group difference testing problem. We provide the testing procedure as well. Fourth, we find that LSTM network not only helps store the long-term dependency to enhance out-of-sample prediction but also keeps the smallest out-of-sample performance decay compared to the benchmark models.
Finally, we implement the arbitrage portfolio with its K-Means clustering to provide an economic interpretation for Benjamin Graham's quote about the market as a short-term ``voting machine" and as long-term ``weighing machine".


The rest of this paper is arranged as follows: In Section \ref{sec:literature}, we summarize the related literature to our paper. Section \ref{sec:model} introduces our pseudo-Siamese networks with LSTM for conditional asset pricing models. Section \ref{sec:result} presents our empirical evidence. Section \ref{sec:conclusion} concludes our contribution and findings.

\section{Relationship to Prior Literature} \label{sec:literature}

Empirical asset pricing aims to explain either the cross-section or the time-series of asset returns. Under the setting of classical expected return-beta representation, expected returns of asset $ i $ are determined by time-invariant risk exposure, $ \beta_i$, and risk factors $\lambda_t $. However, a practical limit addresses a challenge that existing research papers can hardly find a good stochastic discount factor (SDF) $ M_{t+1} $, which guarantees the zero mispricing errors, $ \alpha_i $ as possible. The issue of nonzero alpha has been widely documented in the asset pricing literature of the factor models \citep{nagel_empirical_2013,fama_five-factor_2015, fama_international_2017, hou_digesting_2015}. For the conditional asset pricing models in this paper, the idea of dynamic spanning of risk exposures and mispricing errors, i.e. $\beta_{it} $ and $ \alpha_{it}$, is more closely related to the real market.

In the nearly 28 years following the Fama-French 3 factor model \citep{fama_cross-section_1992}, hundreds of asset pricing papers attempt to capture and identify factor exposures which are unable to be explained by the 3-factor model. To address this problem, researchers have succeeded in forming characteristics-based factor models to predict the cross section of stock returns \citep{daniel_evidence_1997, chen_characteristics-based_2020}. In particular, in a detailed work of ``factors fishing", \cite{harvey__2016} provide a multiple testing for 316 proposed factors over 313 published papers that study the cross-section of expected returns. Given their recommended test thresholds, a $ t $-statistic of 3.0, they argue that more than 100 of factors are false discoveries, leaving nearly 130 characteristics as useful predictor for stock returns. \cite{freyberger_dissecting_2020},  contribute a seminal paper on identifying which firm-specific characteristics of 62 given characteristics carry incremental information. \cite{kim_arbitrage_2020} use Projected-PCA to form a characteristics-based alpha and beta model, which further demonstrates that, indeed, firm characteristics can be used for explaining variation of stock returns via linking to alpha and beta. Following the directions of these research, we choose 90 firm characteristics as input features for the deep beta and deep alpha network, respectively, to capture the firm-specific information.

In addition, as found by recent research papers, \cite{lettau_resurrecting_2001, lettau_factors_2019, chen_deep_2020}, including macroeconomic information in the asset pricing model plays an important role in explaining variation of asset prices and improves out-of-sample forecasting. To describe the fundamental macroeconomic sources of risk in our conditional asset pricing models, we also incorporate 124 macroeconomic variables and market excess returns as the partial input features for our risk factors network.

The high dimensional information of financial characteristics and macroeconomic states raises the question of how to handle the curse of dimensionality. A burgeoning literature of machine learning in empirical asset pricing \citep{rapach_international_2013, feng_deep_2018, kelly_characteristics_2019,choi_alpha_2020,gu_empirical_2020, chen_deep_2020, kozak_shrinking_2020}, shows how machine learning tools are capable of solving this issue. However, \cite{avramov_machine_2020} shows that the machine learning-based return signal is down dramatically after imposing some economic restrictions, such as excluding microcaps. To tackle with the absence of economic interpretation, \cite{cong_alphaportfolio_2020} incorporate the economic distillation through linear projection, while such a statistical manipulation is not directly linked to any economic theory. Distinct from the existing asset pricing literature on machine learning, we apply the Siamese network under the classical single-beta representation model and allow for an adversarial relationship among alphas, betas, and risk factors in explaining variation of stock returns. Such configuration builds a bridge between deep learning approaches and asset pricing theories, which emphasizes the important role of economic foundation in implementing deep learning methods. Moreover, because the Siamese network is an effective method in learning a relationship between two or more comparable items discriminatively \citep{chopra_learning_2005}. In addition, the LSTM  as the sub-network in the Siamese network, allows the sub-network to retain long-term dependencies of input features at a given time $ t $ from many timestamps before, which helps store long-term financial information or macroeconomic sources for estimating the deep alphas, deep betas and risk factors. 

\cite{ferson_conditioning_1999} and \cite{kim_arbitrage_2020} construct an arbitrage portfolio to assess the economic meaning of the mispricing errors. Based on their work, we further dig into the deep detail of the mispricing errors. In particular, we use K-Means algorithm to classify the deep alphas for forming different clusters of arbitrage portfolios. Instead of using portfolio returns, we evaluate the arbitrage portfolio with its K-Means clustering through the risk-adjusted returns, the Sharpe ratio. We investigate the time series properties of arbitrage portfolio via its Sharpe ratio for different clustering and verify the quote of Benjamin Graham based on that.

\section{Model} \label{sec:model}

%
In this section, we describe asset returns in terms of conditional moments and how we configure them in a deep learning framework. We start with a general form of a nonlinear stochastic discount factor of factor $ f_{t+1} $, $ M_{t+1} = d(f_{t+1}) $, where $ d(\cdot) $ is differentiable and $ |E_{t}(f_{t+1})| < \infty $ with probability 1\footnote{$ M_{t+1} $ is approximated under Taylor expansion. A special case shown by \textit{Stein's lemma} is the approximated symbol can be replaced by the equal sign when factors $ f_{t+1} $ and returns $ R_{t+1} $ are bivariate normal.}. As shown below, in spite of a nonlinear stochastic discount factor, it can still be simplified into the linear form of factors models. 
\begin{align*}
M_{t+1} & = d(f_{t+1}) \approx d(E_{t}(f_{t+1})) + d'(E_{t}(f_{t+1}))(f_{t+1} - E_{t}(f_{t+1})) \\
& = \underbrace{d(E_{t}(f_{t+1})) - d'(E_{t}(f_{t+1}))E_{t}(f_{t+1})}_{a_t} +  \underbrace{d'(E_{t}(f_{t+1}))}_{b_t}f_{t+1}  \\
& = a_t + b_t f_{t+1}
\end{align*}
Note that the basic asset pricing equation $ 0 = E_t[M_{t+1}R_{i,t+1}^e]  $ does not identify $  E_t[M_{t+1}]$, so here we can normalize $ a_t $ arbitrarily. It is convenient to normalize the linear representation of stochastic discount factor to $ M_{t+1}  =  1 + b_{t}[f_{t+1}  - E_{t}(f_{t+1})] = 1 + b_t \tilde{f}_{t+1}$ with $ \tilde{f}_{t+1} =  f_{t+1}  - E_{t}(f_{t+1})$.

According to the equivalence theorem between factor models and discount factors, we obtain the single-beta representation as shown in \autoref{eq:onebeta} \footnote{The equivalence theorem between factor models and discount factors addresses that given any multiple factors model, one can find a single-beta representation, that implies that the number of factors is not a meaningful question \citep{cochrane_asset_2009}.}.
\begin{align}
0 = E_t[M_{t+1}R_{i,t+1}^e]    \iff  E_t[R_{i,t+1}^e] = \underbrace{  \frac{cov_{t}(\tilde{f}_{t+1}, R_{i,t+1}^e)}{Var_{t}(\tilde{f}_{t+1})}}_{\beta_{it}(E_{t}(f_{t+1}), E_{t}(f^2_{t+1}),E_{t}(R^e_{i,t+1}))} \underbrace{- b_{t}Var_{t}(\tilde{f}_{t+1})}_{\lambda_{t}(E_{t}(f_{t+1}),E_{t}(f^2_{t+1}))}
\label{eq:onebeta}
\end{align}
where $  R_{i,t+1}^e $ denotes equity returns in excess of risk-free rate by $  R_{i,t+1} - R_{t}^f $ . Notice that $ \beta_{it}$ is formed as a nonlinear and complex function of $ E_{t}(f_{t+1} )$, $ E_{t}(f^2_{t+1}) $, and $ E_{t}(R^e_{i,t+1}) $ and so is $  \lambda_{t} $ (function of $ E_{t}(f_{t+1}) $ and $ E_{t}(f^2_{t+1}) $ only).
In addition, when taking the unconditional expectation of \autoref{eq:onebeta}, the unconditional asset pricing model
\begin{align}
E[R_{i,t+1}^e]  = cov(\beta_{it}, \lambda_t) + E(\beta_{it})E(\lambda_{t})  \stackrel{?}{=} E(\beta_{it})E(\lambda_{t})
\label{eq:uncondition}
\end{align}
only fulfills if the additional covariance term, $ cov(\beta_{it}, \lambda_t) $, is zero. However, as shown above,  $ \beta_{it}$ and $ \lambda_{t} $ are both formed by $ E_{t}(f_{t+1}) $ and $ E_{t}(f^2_{t+1}) $, which implies that the unconditional covariance term, $ cov(\beta_{it}, \lambda_t) $, will not generally be $ 0 $. Therefore, traditional tests of empirical asset pricing models based on the unconditional model, including CAPM and multi-factors models, are misspecified when risk exposure and factor risk premia are correlated and time-varying. One the other hand, \cite{aymanns_models_2018} point out that many investors take alpha very seriously, as evidenced by overlapping portfolios across financial networks. The network of mispricing can facilitate risk diversification and reduce the overlap across portfolios. So we also take into account time-varying alphas in \autoref{eq:onebeta} to gauge asset mispricing.

The partial solution of testing conditional pricing models is to create scaled factors, $ f_{t+1}\otimes z_t $ with conditioning variables, $ z_t $, such as firm characteristics \citep{cochrane_cross-sectional_1996, hodrick_evaluating_2001}. This method strictly requires that conditioning variables be parsimoniously selected and capture the information set regarding future excess returns and investor the information set in a linear way. Because the partial solution uses limited variables in a linear way, we can see that such a partial solution fails to handle high dimensional information set with complex function as shown in \autoref{eq:onebeta}. Especially, \cite{cochrane_presidential_2011} conjectures ``to address these questions in the zoo of new variables, I suspect we will have to use different methods."

This unsolved question motivates our efforts to tie mispricing, risk exposure, and risk factor premia to a general function of high dimensional conditional information using deep learning approaches. Therefore, we formulate the deep alphas, deep betas, and the deep risk factors from the conditional information set(s) (referred to as input features here) as follows: 
\begin{align}\label{eq:deep}
  \alpha(\mathbf{z_{it}}) = \alpha(\mathbf{z_{it}};  \mathbf{p_{\alpha}}), \alpha(\cdot;  \mathbf{p_{\alpha}}) = \mathcal{F}(L, \mathbf{p_{\alpha}}) := \big\{\text{the form of LSTM in (\ref{eq: lstm}} ) \big\}   \\ \nonumber
  \beta(\mathbf{z_{it}}) = \beta(\mathbf{z_{it}}; \mathbf{p_{\beta}}), \beta(\cdot; \mathbf{p_{\beta}}) = \mathcal{F}(L, \mathbf{p_{\beta}}) := \big\{\text{the form of LSTM in (\ref{eq: lstm})}\big\} \\
  \lambda (\bar{\mathbf{z}}_t, \mathbf{m}_{t}) = \lambda(\bar{\mathbf{z}}_t, \mathbf{m}_{t}; \mathbf{p_{\lambda}}), \lambda(\cdot; \mathbf{p_{\lambda}}) = \mathcal{F}(L, \mathbf{p_{\lambda}}) := \big\{\text{the form of LSTM in (\ref{eq: lstm})}\big\} \nonumber
\end{align} 
where the input features contain $ K $ selected firm characteristics, $ \mathbf{z_{it}} = (z_{i1t},\dotsc,z_{iKt}) $ for $ alpha $ and $ beta $ sub-network individually. $\alpha_{it}$  and $\beta_{it} $ are denoted by $\alpha(\mathbf{z_{it}})$  and $\beta(\mathbf{z_{it}})$, respectively. As risk factors capture the tendency of all assets to vary together, irrespective of their firm characteristics,  we use the average firm characteristics across all characteristics, $ \bar{\mathbf{z}}_t  = (\frac{1}{N_t}\sum_{i=1}^{N_t}z_{i1t},\dotsc, \frac{1}{N_t}\sum_{i=1}^{N_t}z_{iKt})$, and $ J $ macroeconomic states, $\mathbf{m}_{t} = (m_{1t},\dotsc, m_{Jt})  $ as input features for $ \lambda $ sub-network. $  \mathcal{F}(L, \mathbf{p})$ defines the space of network functions with given the number of hidden layers, $ L $ and a width parameter vector $ \mathbf{p} = (p_0, \dotsc, p_{L+1})$. Here we assume that pricing information is persistent and can predict the future stock returns.

Plugging \autoref{eq:deep} into (\ref{eq:onebeta}), therefore, we obtain a one-factor model with the deep risk factors, deep betas, and deep alphas as follows:
\begin{align}
	R_{i,t+1}^e = 	 \underbrace{\alpha(\mathbf{z}_{it}) +	\beta(\mathbf{z}_{it}) \lambda(\bar{\mathbf{z}}_{t}, \mathbf{m}_{t})}_{pseudo-Siamese\, network} + \epsilon_{i,t+1}
\label{eq:model1}
\end{align} 
where $ E[\epsilon_{i,t+1}|\mathbf{z_{it}},\mathbf{m}_{t}]  = 0$. The conventional two-pass estimation method, i.e. estimating \autoref{eq:model1} first time-series and then cross-sectional regressions \citep{cochrane_asset_2009}, is invalid here because of variations of alphas and betas over time and complex functions of high dimensional variables specified in \autoref{eq:model1}. Estimating \autoref{eq:model1} also helps us solve the above questions through the following methods: (i)  identification of the cross-section and time series variation of realized asset returns; (ii) decomposition of asset returns into predictable systematic components and unpredictable non-systematic components.
\subsection{pseudo-Siamese Network for Asset Pricing (SNAP)}
\begin{figure}[!htbp]
	\caption{pseudo-Siamese Network for Asset Pricing (SNAP) \label{fig:cs_siamese}}
	\includegraphics[width=1\textwidth]{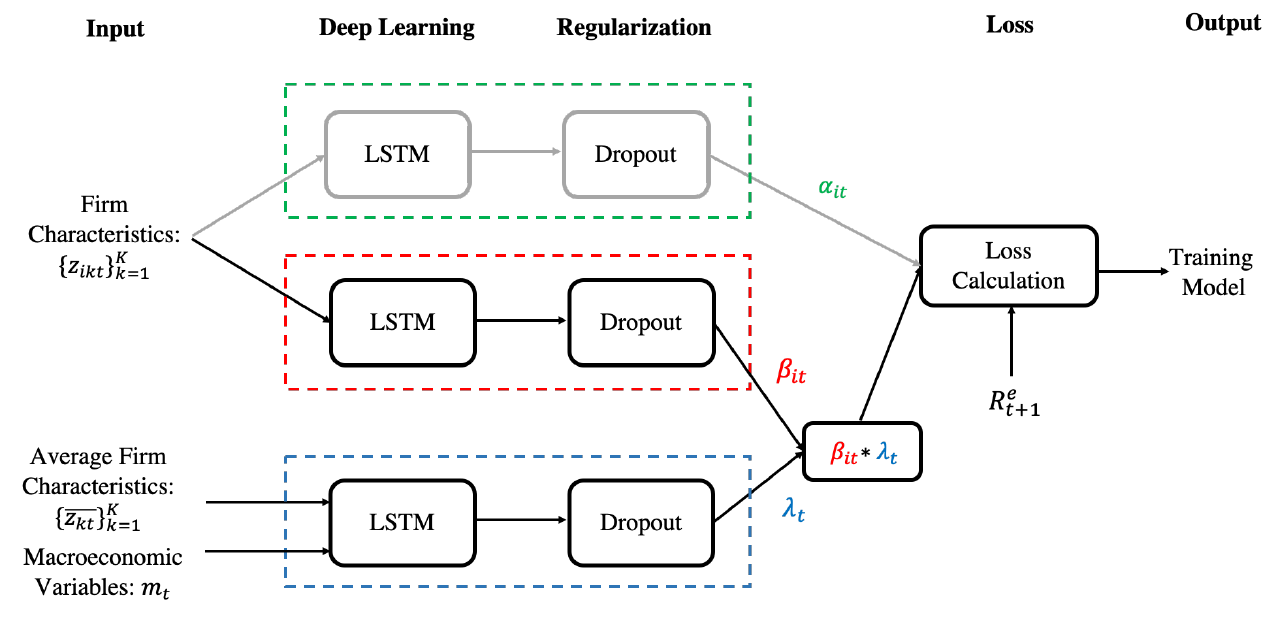}
	\footnotesize Note: \autoref{fig:cs_siamese} is the block diagram of the pseudo-Siamese Network for Asset Pricing (SNAP). This figure illustrates the model architecture of SNAP with LSTM networks. The overall model involves three major branches, i.e., mispricing $ \alpha_{it} $, risk exposure $ \beta_{it}$, and risk factors $ \lambda_{t} $. For $ \alpha_{it} $ branch and $ \beta_{it}$ branch, 90 firm characteristics as inputs, dynamic latent features are extracted by LSTM sub-network and regularized by dropout for final training. Similar procedure applied to $ \lambda_{t} $ branch, however, its inputs contain total 215 features, including 90 average firm characteristics and 125 macroeconomic states (124 macroeconomic variables and market excess returns). And $ \beta_{it}$ branch and $ \lambda_{t} $ branch are concatenated through multiplication as shown in \autoref{eq:model1}. Dropout is applied to each LSTM network for reducing overfitting and improving model performance.
\end{figure}

We use a pseudo-Siamese network to integrate with all of three pricing components and excess returns from the conditional asset pricing models through one network. Siamese neural network is a neural network architecture containing two or more identical sub-networks in which the sub-networks are configured with the same parameters and weights \citep{bromley_signature_1994}. However, as shown in \autoref{eq:model1}, either risk exposure or mispricing carries the firm-specific information while factor risk premia involves the common information set for each individual firm. Therefore, we make three different branches of the pseudo-Siamese network which provides more flexibility but requires more computational power. The pseudo-Siamese network for the conditional asset pricing models described above is as depicted in \autoref{fig:cs_siamese}.

\autoref{fig:cs_siamese} visualizes the pseudo-SNAP model, illustrating the connection among the deep alphas, deep betas, and deep risk factor in a one-pass framework, which classifies the effects among these three pricing elements to asset returns in an adversarial manner. Empirically, we find that when $ \alpha_{it} $  and $ \beta_{it} $ share the same tuning parameters, our pseudo-SNAP models achieve  better performance. Therefore, our empirical results focus on the case of $ \alpha_{it} $  and $ \beta_{it} $ sharing the same tuning parameters.


The following properties of the pseudo-SNAP model explain why our model matters: 1) \textit{flexible functional form for each pricing element and high dimensional information sets}: Each pricing element is able to be independently represented by a nonlinear and flexible network function, such as convolutional neural network (CNN), recurrent neural network (RNN), autoencoder, LSTM, etc. In other words, the pseudo-SNAP model does not depend on any stringent assumptions about the function form of input features and is able to adapt any scenario. More importantly, model flexibility enhances predictions; We also allow each pricing element to be a general network function of high dimensional conditional information sets. 2) \textit{prediction refinement}: In each training step, the pseudo-Siamese network helps judge how similar the excess returns predictor is to the observable excess returns and refines the out-of-sample predictive ability from the previous training step via the Adam algorithm. 3) \textit{long-term dependencies}: The very deep neural network may suffer from the vanishing and exploding gradient problem. Hence, we implement the long short-term memory network in our pseudo-Siamese architecture, which stores the long-run memory of the sequence of firm characteristics and macroeconomic variables to further improve the out-of-sample predictive performance of our model. 4) \textit{regularization}: We use Dropout to avoid the problem of overfitting.
The empirical loss function of the pseudo-SNAP model minimizes the average sum of squared prediction errors as below:
\begin{align}
	\min_{\mathbf{p_{\alpha}}, \mathbf{p_{\beta}}, \mathbf{p_{\lambda}}} \frac{1}{N}\sum_{i=1}^{N}\frac{1}{T_i}\sum_{t=1}^{T_i} [R^e_{i,t+1} -[\underbrace{\alpha(\mathbf{z}_{it}) +	\beta(\mathbf{z}_{it}) \lambda(\bar{\mathbf{z}}_{t}, \mathbf{m}_{t})}_{pseudo-Siamese\, network}]]^2
	\label{eq:loss}
\end{align}
where  $ \mathbf{p_{\alpha}}= (W_{1\alpha}, W_{2\alpha}, B_{\alpha} ) $, $ \mathbf{p_{\beta}} = (W_{1\beta},W_{2\beta},B_{\beta}) $, $ \mathbf{p_{\lambda}} = (W_{1\lambda} , W_{2\lambda},B_{\lambda}) $ denotes the tuple of the weight matrices and bias vectors. Given such a large set of parameters and non-linearity of neural networks, the conventional brute force optimization is computationally inefficient to optimize the mode. In terms of optimization algorithms, we choose Adam as the optimizer, which not only solves for non-convex problems but also is appropriate for non-stationary objects and sparse gradients problems \citep{kingma_adam_2017}.   The algorithm of our pseudo-SNAP algorithm with LSTM network is summarized in Algorithm, the pseudo-SNAP model as below.
\begin{table}[!htbp]
	\caption{Algorithm of the pseudo-Siamese Network for Asset Pricing (SNAP) model}
	   \begin{tabular*}{1\columnwidth}{l}
    	\toprule
    \textbf{Algorithm}:  The pseudo-\textbf{SNAP} model \\
    \midrule
    \textbf{Input}: \\
    \qquad  Normalize the input features, $ \mathbf{z}_{it} $, $ \bar{\mathbf{z}}_{t} $, and $ \mathbf{m}_{t}$\\
   \textbf{Training}: \\
   \qquad  Initialize the tuning parameters: \\
   \qquad \quad   \#\textit{epochs}, \textit{batch size}, \#\textit{hidden neurons}, \textit{dropout rate}, \textit{learning rate} \\
    \qquad \textbf{for} $ i = 1 $ to $ N $ \textbf{do}  \\
     \qquad \qquad \textbf{for} $ t = 1 $ to $ T_i $ \textbf{do}  \\
     \qquad \qquad  \qquad Feed the individual input, $ \mathbf{z}_{it} $ into $ \alpha_{it} $, $ \beta_{it} $ LSTM network using eq. (\ref{eq:model1}) \\
      \qquad \qquad  \qquad Feed the common features, $ \bar{\mathbf{z}}_{t} $ and $ \mathbf{m}_{t}$ into $ \lambda_{t} $ LSTM  network using eq. (\ref{eq:model1}) \\
   \qquad \qquad  \qquad  Obtain the $ \alpha_{it}$, $\beta_{it}$, and $\lambda_{t}$ LSTM prediction results \\
    \qquad \qquad  \qquad Concatenate three branches via a pseudo-Siamese network \\
    \qquad \qquad  \textbf{end for}  \\
     \qquad  \textbf{end for}  \\
    Calculate the prediction of excess returns, $ \hat{R^e}_{i,t+1} $, \\
    Update parameters by optimizing the loss function in eq. (\ref{eq:loss}) through the Adam algorithm.  \\
    \bottomrule
    \end{tabular*}%

  \label{tab:snap_algo}%
  \footnotesize Note: This table summarizes the algorithm for our pseudo-SNAP model, including normalization, initialization, training, concatenation, and optimization. 
\end{table}%
\subsection{Regularized long short-term memory (LSTM)}
%
%
The long short-term memory (LSTM), one type of the gated recurrent neural networks (RNNs), achieves state-of-the-art forecasting performance on important tasks \footnote{Such as language modeling, machine translation, and speech recognition.} and has been implemented in many popular virtual assistants \footnote{As in Apple's Siri, Google's Voice Search, and Samsung's S Voice.}, including economic and financial time series \citep{ smalter_hall_macroeconomic_2017, siami-namini_forecasting_2018}. We implement the LSTM network to our input features, firm characteristics and macroeconomic states, to generate the time varying $ \alpha_{it} $, $ \beta_{it}$, and $ \lambda_{t} $ for the following reasons.

First, as most of financial ratios and macroeconomic variables are non-stationary, the traditional way of handling non-stationarity is taking the difference or de-trending the sequence while a flaw of this method is that it may lead to loss of predictive signal. Second, the long-run dynamics of financial activities and the macroeconomy are important for prediction according to the recent literature. \cite{schularick_credit_2012, jorda_rate_2019} investigate the long-run dynamics of monetary finance, financial markets, and the macroeconomy through an extensive long-history database. They discover that business cycles can certainly affect the economy, especially the financial markets, and emphasize the recurrent episodes of financial instability. \cite{jha_does_2020} propose a sentiment index toward finance to predict economic growth on a long-history dataset and further demonstrate long-term dependency of financial variables and macroeconomic states. The LSTM is capable of filtering financial and macroeconomic regimes via its ``gates" compared to recurrent neural networks, which allows our model to capture structural changes in the series that reoccur over time. Furthermore, with the memory advantage, it generates the similar results in financial products as the multilayer perceptron (MLP, also referred to as deep feedforward networks (FFN)) does \citep{nunes_memory_2019}.  Last, LSTM’s output is only dependent on the previous and current states, which fulfills the requirement of conditional asset pricing models. 

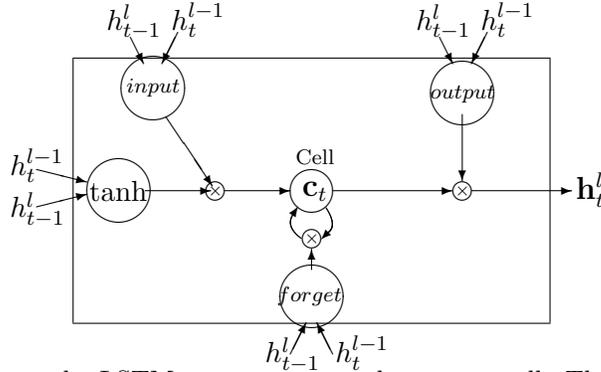
\begin{figure}[!htbp]
		\caption{LSTM architecture indicated via directed acyclic graphs.}
	\begin{center}
		\begin{picture}(200, 130)
			\put(0, 0){\framebox(180, 100){}}
			\put(90, 50){\circle{16}}
			\put(86.5, 48){$\mathbf c_t$}
			\put(84, 60){{\scriptsize Cell}}
			
			\put(90, 32){\circle{6.5}}
			\put(87.25, 30.5){{\tiny $\times$}}
			
			\put(90, 10){\circle{22}}
			\put(77, 8){{\scriptsize $forget$}}

			\put(90, 20){\vector(0, 0){8}}
			
			\put(85, 44){\vector(2, 3){0}}
			\qbezier(86, 32)(80, 36.5)(83, 41)
			
			\qbezier(96, 34)(100, 36.5)(95.5, 43.5)
			\put(93.5, 31.5){\vector(-1, -1){0}}
			
			\put(82, -12){\vector(1, 2){6}}
			\put(72.5, -15){{\small $h_{t-1}^{l}$}}
			\put(98, -12){\vector(-1, 2){6}}
			\put(99, -15){{\small $h_{t}^{l-1}$}}
			
			\put(30, 89){\circle{22}}
			\put(20, 87){{\scriptsize $input$}}

			\put(21.5, 108){\vector(1, -2){5}}
			\put(12.5, 111){{\small $h_{t-1}^{l}$}}
			\put(39.5, 110){\vector(-1, -2){6}}
			\put(37, 111){{\small $h_{t}^{l-1}$}}
			
			\put(147, 87){\circle{22}}
			\put(135, 85){{\scriptsize $output$}}

			\put(138.5, 108){\vector(1, -2){5}}
			\put(129.5, 111){{\small $h_{t-1}^{l}$}}
			\put(156.5, 110){\vector(-1, -2){6}}
			\put(154, 111){{\small $h_{t}^{l-1}$}}
			
			\put(17, 50){\circle{22}}
			\put(6, 46){{\small tanh}}

			\put(53.5, 50){\circle{6.5}}
			\put(50.75, 48.5){{\tiny $\times$}}
			
			\put(57, 50){\vector(1, 0){25}}
			\put(27, 50){\vector(1, 0){25}}
			\put(35, 78){\vector(2, -3){17.5}}
			
			\put(147, 50){\circle{6.5}}
			\put(144.25, 48.5){{\tiny $\times$}}
			\put(98, 50){\vector(1, 0){45.25}}
			\put(150.5, 50){\vector(1, 0){38}}
			
			\put(147, 79){\vector(0, -1){25.5}}
			\put(190, 47){${\mathbf h^l_t}$}
			
			\put(-24, 40){{\small $h_{t-1}^{l}$}}
			\put(-24, 56){{\small $h_{t}^{l-1}$}}
			\put(-14, 44){\vector(4, 1){19}}
			\put(-14, 58){\vector(4, -1){19}}		
		\end{picture}
	\end{center}
	\label{fig:lstm}
	\footnotesize Note: This figure illustrates the LSTM recurrent network memory cell. The cells are linked to each other and three gates, a \textit{forget} gate, \textit{input} gate, and \textit{output} gate regulate information flow in each cell. All the gating units have a nonlinear activation function, we use rectified linear unit (ReLU) activation function. 
\end{figure}

For simplicity of illustration, the regularized LSTM transition equations are defined as follows:
\begin{align}	\label{eq: lstm}
	&\text{LSTM} : h^{l-1}_t, h^l_{t-1}, c^l_{t - 1} \rightarrow h^l_t, c^l_t\\
	&\begin{pmatrix}input\\forget\\output\\g\end{pmatrix} =
	\begin{pmatrix}\mathrm{\sigma}\\\mathrm{\sigma}\\\mathrm{\sigma}\\\tanh\end{pmatrix}
	T_{2d,4d}\begin{pmatrix}{\bf D}(h^{l - 1}_t)\\h^l_{t-1}\end{pmatrix}\\ \nonumber
	&c^l_t = forget \odot c^l_{t-1} + input \odot g\\ \nonumber
	&h^l_t = output \odot \tanh(c^l_t)
\end{align}
where $ h^{l}_{t} $ denotes that all of states are $ d $-dimensional in layer $ l $ at timestamp $ t $. Also, $ T_{d_1,d_2}: \mathbb{R}^{d_1}  \rightarrow  \mathbb{R}^{d_2}$ represents an affine transform function with the format of $( W'x + b) $ for input $ x $ given the weight matrices $ W $ and  bias term $ b $.
The key to LSTM is the memory cell that has the function to add or remove information from previous and current states through three gates, \textit{forget} gate, \textit{input} gate, and \textit{output} gate \citep{graves_generating_2014}. Figure \ref{fig:lstm} summarizes the detailed architecture of the memory cell.
In other words, if the time series is long and the previous information is far but useful, LSTM is able to store it for current state, while other RNNs would suffer from the gradient exploding or decaying exponentially \citep{bengio_learning_1994}.
The non-linear transformation, $ \sigma $, triggered in each gate is by the rectified linear unit (ReLU) activation function, i.e. $ \sigma = ReLU(x) = max(x,0) $ \footnote{Other widely used nonlinear activation functions include the sigmoid (also referred to as the logistic function) and the tanh (also referred to as the hyperbolic tangent function). However, these functions are suffering the limitation of sensitivity and saturation \citep{goodfellow_deep_2016}, for instance, vanishing gradients issue.} and the memory cell $ c_t^l $ is activated by the tanh for $ h_t^l $. $ L $ denotes the number of layers in LSTM and we use $ h^L_t $ to predict our output.


As many large neural networks with a large number of parameters are hard to tackle with overfitting, dropout is a regularization method for addressing this issue through randomly dropping units from the network during training \citep{srivastava_dropout_2014}. In particular, \cite{zaremba_recurrent_2015} find that the recipe for applying dropout to LSTM is that the dropout operator, $ \mathbf{D} $ is only applied to the non-recurrent part as shown in \autoref{eq: lstm}.

\subsection{Mispricing: pseudo-SNAP v.s. masked pseudo-SNAP}
\begin{figure}[!htbp]
	\caption{The masked $ \alpha_{it} $ pseudo-SNAP for mispricing testing \label{fig:masked_siamese}}
	\includegraphics[width=1\textwidth]{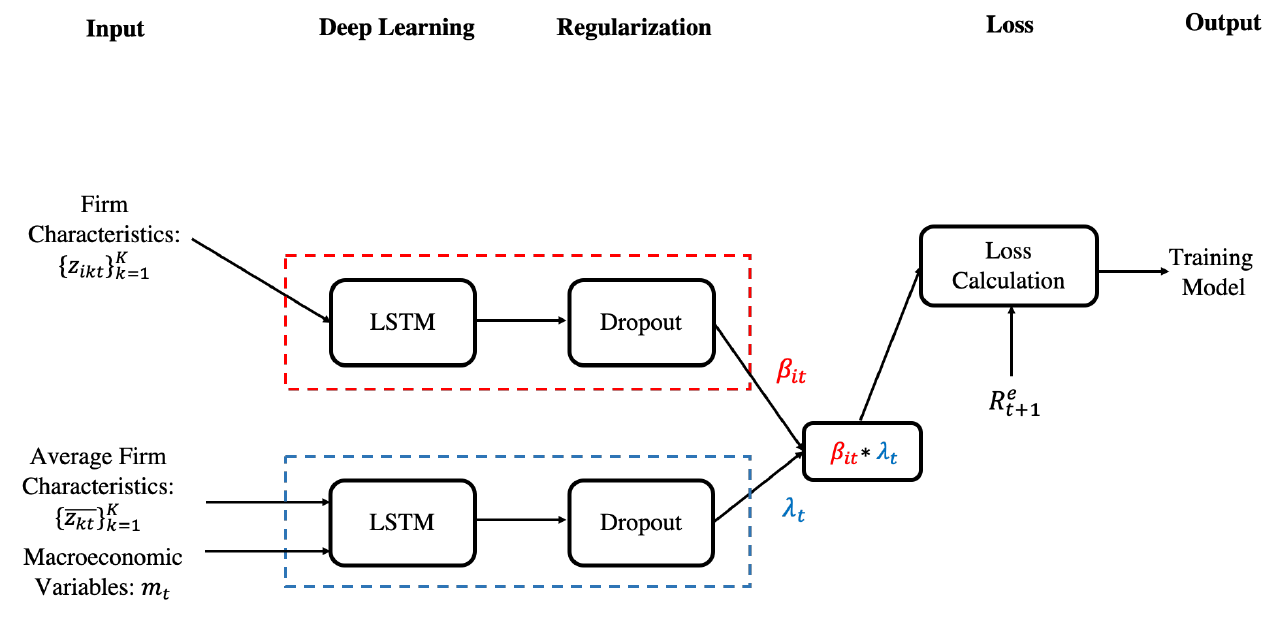}
	\footnotesize Note: \autoref{fig:masked_siamese} is the block diagram of the masked pseudo-Siamese Network for Asset Pricing (SNAP). This figure illustrates the $ \alpha_{it} $ sub-network is masked from the original pseudo-SNAP model in \autoref{fig:cs_siamese}. In other words, the overall model only contains two branches, i.e., risk exposure $ \beta_{it}$ and risk factors $ \lambda_{t} $. To evaluate the pseudo-SNAP model and test for zero mean pricing errors, we develop a t statistics test or Mann-Whitney test for the two group residuals, one is obtained from the pseudo-SNAP model, the other one is calculated from the masked $ \alpha_{it} $ sub-network one.
	\label{fig:snap2}
\end{figure}

An important question emerges from \autoref{eq:model1} is whether or not to include $ \alpha_{it} $ branch in our pseudo-SNAP model, which is equivalent to testing for $ H_0: \alpha_{it} = 0 $ and to evaluating our pseudo-SNAP model.  In doing so, we disentangle the mispricing hypothesis testing into a controlled experiment question. We first treat the original pseudo-SNAP model as a control group and mask out the deep alphas sub-network only with other sub-network held constant as an experimental group. Masking out the $ \alpha_{it} $ branch from \autoref{fig:cs_siamese} and obtaining the masked pseudo-SNAP model are shown in \autoref{eq:model2} and \autoref{fig:snap2}.
\begin{align}
	[R_{i,t+1}^e]^{masked}_{\alpha_{it} = 0}= 		\beta(\mathbf{z}_{it}) \lambda(\bar{\mathbf{z}}_{t}, \mathbf{m}_{t}) + \epsilon_{i,t+1}
	\label{eq:model2}
\end{align}
Then we calculate the residuals for the unmasked and masked model, respectively, and the estimate of $ \alpha_{it} $ is the difference between these residuals.
\begin{align}
	\label{eq:xestimate}
\hat{\epsilon}_{i,t+1} & = R^e_{i, t+1} -  \hat{R}^e_{i, t+1} \\ \nonumber
\hat{\epsilon}^{masked}_{i,t+1} & = R^e_{i, t+1} - [ \hat{R}^{e}_{i, t+1}]^{masked}_{\alpha_{it} = 0} \\
\hat{\alpha}_{it} & =  \hat{\epsilon}^{masked}_{i,t+1}  -  \hat{\epsilon}_{i,t+1}  \nonumber 
\end{align}
where $ \hat{R}^e_{i, t+1} $ and $[ \hat{R}^{e}_{i, t+1}]^{masked}_{\alpha_{it} = 0}  $ denotes as prediction value estimated from the pseudo-SNAP model and masked pseudo-SNAP model, respectively, and similar definitions are used for $ \hat{\epsilon}_{i,t+1} $ and $ \hat{\epsilon}^{masked}_{i,t+1} $. To make the estimate of $ \alpha_{it} $ consistent, we use the same tuning parameters for both models except the $ \alpha_{it} $ network. As $ \alpha_{it} $ is in unbalanced panel data format, we focus on the testing for $ H_0: \alpha_{it} = 0 $ in an unconditional perspective, which is equivalent to testing for the classical group mean difference between these two residuals, defined as below:
\begin{align}
H_0: \mu(\hat{\epsilon}^{masked}_{i,t+1})  = \mu(\hat{\epsilon}_{i,t+1} ) \quad v.s. \quad H_1:  \mu(\hat{\epsilon}^{masked}_{i,t+1}) \neq \mu(\hat{\epsilon}_{i,t+1})
\label{eq:alpha_test}
\end{align}
It is worth noting that the traditional $ t $ test statistic for the group difference must fulfill with the assumption that both residuals are sampled from normal distributions with equal variances. Such a strong assumption makes $ t $ test contingent in this scenario. Therefore, we design a mispricing (alpha) testing procedure as follows: 1) Check normality of both residuals using either Shapiro-Wilk test or Kolmogorov-Smirnov test; 2) If the normality test rejects both residuals are normal, we use the Mann-Whitney $ U $ test for testing mispricing instead while the $ t $ test statistic would be used if the normality test is accepted.
\subsection{Economic interpretation: K-Means Arbitrage portfolios}
Following the recent work of \cite{kim_arbitrage_2020}, each month we construct arbitrage portfolios using the estimate of mispricing (alphas) in \autoref{eq:xestimate}. The return of the arbitrage portfolios is defined as $ R^{ap}_{t+1} = \mathbf{w}_t'\mathbf{R}_{t+1}^e $, where $ \mathbf{w}_t = \hat{\mathbf{\alpha}}_t/N_t $. However, this aggregated construction ignores the detailed network behind the mispricing information \citep{li_dynamic_2020}. For instance, instead of constructing the portfolios based on all of the stocks, some specialists in the trading firms or the funds, are capable of achieving higher performance by picking more outstanding stocks only. One conjecture is that these specialists are able to generate the large alphas in their diversified portfolios. To investigate this conjecture, we utilize the K-Means clustering method to classify different groups of arbitrage portfolio in every month\footnote{The K-Means algorithm is a popular iterative descent clustering methods\citep{cox_note_1957}.}. Then we select the highest, median, lowest cluster of arbitrage portfolio according its Sharpe ratio. Such a classification leads us to find the unknown data-driven pattern of mispricing information. The algorithm of  K-Means clustering of the arbitrage portfolios is summarized in \autoref{tab:kmeans_algo}.
\begin{table}[!htbp]
	\caption{Algorithm of arbitrage portfolio K-Means clustering}
	   \begin{tabular*}{1\columnwidth}{l}
    	\toprule
    \textbf{Algorithm}:  Arbitrage portfolio K-Means clustering \\
    \midrule
    \textbf{Input}: \\
    \qquad  Frame the 2-tuple object, $ x_i = (\alpha_i, R^e_i) $\\
    \qquad Use the Elbow method to determine the number of clusters, $ K $\\
   \textbf{Training}: \\
   \qquad \textbf{for} $ t = 1 $ to $ T $ \textbf{do} \\
   \qquad \qquad  1) Given cluster job $ G $, the total cluster variance \\
    \qquad \qquad \qquad $ \min_{G, {m_k}} \sum_{k=1}^{K}N_k \sum_{G(i)=k}||x_i - m_k||^2 $ \\
   \qquad  \qquad  is minimized w.r.t.  the means of clusters $ \bar{x}_S  = \arg\min_{m} \sum_{i\in S}||x_i - m||^2 $ \\
     \qquad  \qquad  2) Given a set of means, the total cluster variance \\
      \qquad \qquad \qquad $ G(i) = \arg\min_{1\leq k \leq K} \sum_{i\in S}||x_i - m_k||^2 $ \\
     \qquad  \qquad is minimized by segmenting each obs to the closest cluster mean.\\
     \qquad \qquad  iterate 1 and 2 until convergence.\\
    \qquad \textbf{end for}\\
    \textbf{Output}:\\
    \quad return the cluster labels within each month \\
    \quad in each month construct an arbitrage portfolio within each cluster \\
    \bottomrule
    \end{tabular*}%

	\label{tab:kmeans_algo}%
\end{table}%
\subsection{Model Comparison}
Typically, the evaluation of different conditional asset pricing models is based on the out-of-sample forecasting and Sharpe ratio for which portfolios are formed by model predictions for stock excess returns. For each model and each data frame, we report the out-of-sample forecasting for one-month forward excess returns $ R^2_{predictive} $ and the annualized Sharpe ratio of a zero-net-investment portfolio, $ SR $ as defined below. In particular, we construct a zero-net-investment portfolio based on sorting stocks into deciles by the out-of-sample prediction, which short sells the 1st decile stocks of the lowest expected excess returns and longs the 10th decile stocks of the highest expected excess returns.
\begin{align*}
	R^2_{predictive} = 1 - \frac{\sum_{(i,t)\in data (\hat{\epsilon}_{i, t+1})^2}}{\sum_{(i,t)\in data (R^e_{i, t+1})^2}} , \quad SR = \frac{E[R_{zero-net,t}]}{SD[R_{zero-net,t}]}
\end{align*}
We compare our pseudo-SNAP model with other popular approaches pervasive in empirical asset pricing literature, including Fama French 3-factor model (FF3), Fama French 5-factor model (FF5), Least Absolute Shrinkage Selector Operator (LASSO), elastic net regression, ridge regression, and Feedforward network (FFN). \footnote{The benchmark models are described in the Appendix.}


\section{Empirical Analysis} \label{sec:result}

\subsection{Data}
We collect monthly equity return data for all stocks from the Center for Research in Security Prices (CRSP), including delisting returns \citep{shumway_delisting_1999} from January 1970 to December 2019. The one-month Treasury bill rates (i.e. the risk-free rate for generating excess returns) are from Kenneth French Data Library.

Following the previous literature in firm characteristics creation, in particular, we select 90 of 102 firm characteristics that have been identified as predictive signals of US equity returns by \cite{green_characteristics_2017}. These characteristics are constructed from CRSP and Compustat database. We begin in 1970 because most of characteristics only become available in that year and the highest missing rate among firm characteristics is under 30\% as shown in \autoref{tab:missing_summary} \footnote{To impute the missing value of characteristics, we focus on using the stocks with full characteristics in that specific sample division, such as training sample. Then we impute a missing value of a characteristic with that characteristic's cross-sectional median in that month}. We follow standard conventions to restrict our data construction to common stocks traded on the NYSE, AMEX, or NASDAQ that have a non-missing value for common equity in their annual financial statements and a month-end market value on CRSP.
The 90 characteristics are listed in \autoref{tab:firm_char}. 
We provide the details of each characteristic and its source in the Appendix. \footnote{Note that the characteristics are from both published and working papers, and publication dates between 1977 and 2016.} Because some firm characteristics' frequency is quarterly or yearly, we recalculate firm characteristics every month. For every specific month $ t $'s return, we match it with firm characteristics available at the end of month $ t-1 $ under the assumption given by \cite{green_characteristics_2017}.
Finally, it leaves us with 20179 stocks and total 2448619 observations.
The summary statistics of our data set of stock-month observations over the period from \sampleperiod is summarized in \autoref{tab:firm_char_summary}.

In order to capture investors' expectations about future business cycle conditions from macroeconomic activities, we have 124 macroeconomic predictors in total which are from the FRED-MD database as defined in \cite{mccracken_fred-md_2016} and $ R_{m}^{e} $, the excess return on the market defined in \cite{fama_cross-section_1992}. Notice that the macroeconomic predictors from \cite{mccracken_fred-md_2016} provide the same predictive content as the Stock and Watson macroeconomic data \citep{stock_james_h_forecasting_2009}. We implement the suggested standard transformation code to have stationary macroeconomic series as suggested in \cite{mccracken_fred-md_2016}.
We split our sample data into three disjoint date periods, which are 36 years of training sample (1970 - 2005), 5 years of validation sample (2006 - 2010), and 9 years of test sample (2011 - 2019). The first training sample is used for training the model given a set of tuning parameters. Based on the training model and given a set of tuning parameters set, we use validation sample to search for the tuning parameters that optimize validation objective function. Finally, we use the test dataset to evaluate our model.
	\begin{landscape}
	\begin{table}[!htbp]
		\caption{Firm characteristics by category}
		\label{tab:firm_char}
		\scriptsize

    \begin{tabular}{*{6}l} 
    \toprule
          & \multicolumn{1}{l}{\textbf{Acronym}} & \multicolumn{1}{l}{\textbf{Firm Characteristic}} &       & \textbf{Acronym} & \textbf{Firm Characteristic} \\
          \midrule
          & \multicolumn{1}{l}{\textbf{\underline{Momentum}}} &       &       & \textbf{\underline{Investment}} &  \\
    (1)   & \multicolumn{1}{l}{\textit{$\Delta$mom}} & \multicolumn{1}{l}{Change in 6-month momentum} & (45)  & \textit{cinvest} & Corporate investment \\
    (2)   & \multicolumn{1}{l}{\textit{indmom}} & \multicolumn{1}{l}{Industry momentum} & (46)  & \textit{$\Delta$capx\_ia} & Industry-adjusted \% change in capital expenditures \\
    (3)   & \multicolumn{1}{l}{\textit{mom1}} & \multicolumn{1}{l}{1-month momentum} & (47)  & \textit{$\Delta$emp\_ia} & Industry-adjusted change in employees \\
    (4)   & \multicolumn{1}{l}{\textit{mom6}} & \multicolumn{1}{l}{6-month momentum} & (48)  & \textit{chinv} & Change in inventory \\
    (5)   & \multicolumn{1}{l}{\textit{mom12}} & \multicolumn{1}{l}{12-month momentum} & (49)  & \textit{$\Delta$shrout} & Change in shares outstanding \\
    (6)   & \multicolumn{1}{l}{\textit{mom36}} & \multicolumn{1}{l}{36-month momentum} & (50)  & \textit{egr} & Growth in common shareholder \\
          & \multicolumn{1}{l}{\textbf{\underline{Trading Frictions}}} &       & (51)  & \textit{grcapx} & Growth in capital expenditures \\
    (7)   & \multicolumn{1}{l}{\textit{aeavol}} & \multicolumn{1}{l}{Abnormal earnings announcement} & (52)  & \textit{invest} & Capital expenditures and inventory \\
    (8)   & \multicolumn{1}{l}{\textit{beta}} & \multicolumn{1}{l}{Market beta} & (53)  & \textit{IPO} & New equity issue \\
    (9)   & \multicolumn{1}{l}{\textit{beta2}} & \multicolumn{1}{l}{Beta squared} & (54)  & \textit{pctacc} & Percent accruals \\
    (10)  & \multicolumn{1}{l}{\textit{divo}} & \multicolumn{1}{l}{Dividend omission} &       & \textbf{\underline{Value and Growth}} &  \\
    (11)  & \multicolumn{1}{l}{\textit{dolvol}} & \multicolumn{1}{l}{Dollar trading volume} & (55)  & \textit{agr} & Asset growth \\
    (12)  & \multicolumn{1}{l}{\textit{idiovol}} & \multicolumn{1}{l}{Idiosyncratic return volatility} & (56)  & \textit{bm} & Book-to-market \\
    (13)  & \multicolumn{1}{l}{\textit{ill}} & \multicolumn{1}{l}{Illiquidity} & (57)  & \textit{bm\_ia} & Industry-adjusted book to market \\
    (14)  & \multicolumn{1}{l}{\textit{maxret}} & \multicolumn{1}{l}{Maximum daily return} & (58)  & \textit{c2d} & Cash flow to debt \\
    (15)  & \multicolumn{1}{l}{\textit{mve}} & \multicolumn{1}{l}{Size} & (59)  & \textit{c2p} & Cash flow to price \\
    (16)  & \multicolumn{1}{l}{\textit{mve\_ia}} & \multicolumn{1}{l}{Industry-adjusted size} & (60)  & \textit{c2p\_ia} & Industry-adjusted cash flow to price \\
    (17)  & \multicolumn{1}{l}{\textit{pdelay}} & \multicolumn{1}{l}{Price delay} & (61)  & \textit{cash} & Cash holdings \\
    (18)  & \multicolumn{1}{l}{\textit{retvol}} & \multicolumn{1}{l}{Return volatility} & (62)  & \textit{cashpr} & Cash productivity \\
    (19)  & \multicolumn{1}{l}{\textit{roavol}} & \multicolumn{1}{l}{Earnings volatility} & (63)  & \textit{convind} & Convertible debt indicator \\
    (20)  & \multicolumn{1}{l}{\textit{rsup}} & \multicolumn{1}{l}{Revenue surprise} & (64)  & \textit{d2p} & Dividend to price \\
    (21)  & \multicolumn{1}{l}{\textit{baspread}} & \multicolumn{1}{l}{Bid-ask spread} & (65)  & \textit{$\Delta$curr} & \% Change in current ratio \\
    (22)  & \multicolumn{1}{l}{\textit{std\_dolvol}} & \multicolumn{1}{l}{Volatility of liquidity (dollar trading volume)} & (66)  & \textit{$\Delta$quick} & \% Change in quick ratio \\
    (23)  & \multicolumn{1}{l}{\textit{std\_turn}} & \multicolumn{1}{l}{Volatility of liquidity (share turnover)} & (67)  & \textit{$\Delta$s2i} & \% Change in sales-to-inventory \\
    (24)  & \multicolumn{1}{l}{\textit{sue}} & \multicolumn{1}{l}{unexpected quarterly earnings} & (68)  & \textit{divi} & Dividend initiation \\
    (25)  & \multicolumn{1}{l}{\textit{turn}} & \multicolumn{1}{l}{Share turnover} & (69)  & \textit{hire} & Employee growth rate \\
    (26)  & \multicolumn{1}{l}{\textit{zerotrade}} & \multicolumn{1}{l}{Zero trading days} & (70)  & \textit{isc} & Industry sales concentration \\
          & \multicolumn{1}{l}{\textbf{\underline{Profitability}}} &       & (71)  & \textit{lev} & Leverage \\
    (27)  & \multicolumn{1}{l}{\textit{$\Delta$ato\_ia}} & \multicolumn{1}{l}{Industry-adjusted asset turnover growth} & (72)  & \textit{lgr} & Growth in long-term debt \\
    (28)  & \multicolumn{1}{l}{\textit{$\Delta$depr}} & \multicolumn{1}{l}{\% Change in depreciation } & (73)  & \textit{orgcap} & Organizational captial \\
    (29)  & \multicolumn{1}{l}{\textit{$\Delta$pm\_ia}} & \multicolumn{1}{l}{Industry-adjusted profit margin growth} & (74)  & \textit{sp} & Sales to price \\
    (30)  & \multicolumn{1}{l}{\textit{$\Delta$tx}} & \multicolumn{1}{l}{Tax expense growth} & (75)  & \textit{securedind} & Secured debt indicator \\
    (31)  & \multicolumn{1}{l}{\textit{depr}} & \multicolumn{1}{l}{Depreciation/PP\&E} & (76)  & \textit{sgr} & Sales growth \\
    (32)  & \multicolumn{1}{l}{\textit{ep}} & \multicolumn{1}{l}{Earnings to price} & (77)  & \textit{sin} & Sin stocks \\
    (33)  & \multicolumn{1}{l}{\textit{ear}} & \multicolumn{1}{l}{Earnings announcement return} & (78)  & \textit{sma} & \% Sales growth minus A/R growth \\
    (34)  & \multicolumn{1}{l}{\textit{gma}} & \multicolumn{1}{l}{Gross profitability} & (79)  & \textit{smi} & \% Sales growth minus inventory growth \\
    (35)  & \multicolumn{1}{l}{\textit{gmms}} & \multicolumn{1}{l}{\% Gross margin growth minus sales} & (80)  & \textit{sms} & \% Sales growth minus SG\&A growth \\
    (36)  & \multicolumn{1}{l}{\textit{grltnoa}} & \multicolumn{1}{l}{Growth in long-term net operating assets} & (81)  & \textit{tang} & Debt capacity/firm tangibility \\
    (37)  & \multicolumn{1}{l}{\textit{ms}} & \multicolumn{1}{l}{Financial statements score} &       & \textbf{\underline{Intangibles}} &  \\
    (38)  & \multicolumn{1}{l}{\textit{nincr}} & \multicolumn{1}{l}{Number of earnings increases} & (82)  & \textit{absacc} & Absolute accruals \\
    (39)  & \multicolumn{1}{l}{\textit{operprof}} & \multicolumn{1}{l}{Operating profitability} & (83)  & \textit{acc} & Working capital accruals \\
    (40)  & \multicolumn{1}{l}{\textit{ps}} & \multicolumn{1}{l}{Financial statements score} & (84)  & \textit{age} & \# years since first Compustat coverage \\
    (41)  & \multicolumn{1}{l}{\textit{roa}} & \multicolumn{1}{l}{Return on assets} & (85)  & \textit{curr} & Current ratio \\
    (42)  & \multicolumn{1}{l}{\textit{roe}} & \multicolumn{1}{l}{Return on equity} & (86)  & \textit{quick} & Quick ratio \\
    (43)  & \multicolumn{1}{l}{\textit{roic}} & \multicolumn{1}{l}{Return on invested capital} & (87)  & \textit{rd} & R\&D increase \\
    (44)  & \multicolumn{1}{l}{\textit{tb}} & \multicolumn{1}{l}{Tax income to book income} & (88)  & \textit{salecash} & Sales to cash \\
          &       &       & (89)  & \textit{saleinv} & Sales to inventory \\
          &       &       & (90)  & \textit{salerec} & Sales to receivables \\
    \bottomrule

    \end{tabular}%

		\scriptsize  Note: This table reports the firm characteristics we use in our empirical study by category. The detail of variable definition is reported in the Appendix. The sample period spans  \sampleperiod
	\end{table}
\end{landscape}

	\begin{table}[!htbp]
	\caption{Summary statistics for firm characteristics by category}
	\label{tab:firm_char_summary}
	\scriptsize 

    \begin{tabular*}{1\columnwidth}{rrrrlrrr}
    \toprule
    \multicolumn{1}{l}{\textbf{Acronym}} & \multicolumn{1}{l}{\textbf{Mean}} & \multicolumn{1}{l}{\textbf{STD}} & \multicolumn{1}{l}{\textbf{Medium}} & \textbf{Acronym} & \multicolumn{1}{l}{\textbf{Mean}} & \multicolumn{1}{l}{\textbf{STD}} & \multicolumn{1}{l}{\textbf{Medium}} \\
    \midrule
    \multicolumn{1}{l}{\textbf{\underline{Momentum}}} &       &       &       & \textbf{\underline{Investment}} &       &       &  \\
    \multicolumn{1}{l}{$\Delta$mom} & 0.000 & 0.544 & -0.002 & cinvest & 0.015 & 3.669 & -0.002 \\
    \multicolumn{1}{l}{indmom} & 0.137 & 0.290 & 0.113 & $\Delta$capx\_ia & 7.162 & 73.971 & -0.281 \\
    \multicolumn{1}{l}{mom12m} & 0.124 & 0.563 & 0.063 & $\Delta$emp\_ia & -0.107 & 0.650 & -0.060 \\
    \multicolumn{1}{l}{mom1m} & 0.010 & 0.155 & 0.000 & chinv & 0.011 & 0.054 & 0.001 \\
    \multicolumn{1}{l}{mom6m} & 0.053 & 0.361 & 0.024 & $\Delta$shrout & 0.101 & 0.302 & 0.007 \\
    \multicolumn{1}{l}{mom36m} & 0.292 & 0.820 & 0.173 & egr   & 0.133 & 0.665 & 0.086 \\
    \multicolumn{1}{l}{\textbf{\underline{Trading Frictions}}} &       &       &       & grcapx & 0.785 & 3.163 & 0.131 \\
    \multicolumn{1}{l}{aeavol} & 0.764 & 1.916 & 0.185 & invest & 0.075 & 0.167 & 0.053 \\
    \multicolumn{1}{l}{beta} & 1.090 & 0.651 & 1.013 & IPO   & 0.080 & 0.271 & 0.000 \\
    \multicolumn{1}{l}{beta2} & 1.616 & 1.823 & 1.030 & pctacc & -0.574 & 5.447 & -0.060 \\
    \multicolumn{1}{l}{divo} & 0.028 & 0.165 & 0.000 & \textbf{\underline{Value and Growth}} &       &       &  \\
    \multicolumn{1}{l}{dolvol} & 11.148 & 2.973 & 10.796 & agr   & 0.150 & 0.411 & 0.084 \\
    \multicolumn{1}{l}{idiovol} & 0.065 & 0.037 & 0.057 & bm    & 0.769 & 0.720 & 0.598 \\
    \multicolumn{1}{l}{ill} & 0.000 & 0.000 & 0.000 & bm\_ia & 22.835 & 715.915 & 0.028 \\
    \multicolumn{1}{l}{maxret} & 0.075 & 0.074 & 0.053 & c2d   & -0.003 & 1.247 & 0.130 \\
    \multicolumn{1}{l}{mve} & 11.764 & 2.263 & 11.621 & c2p   & 0.068 & 1.528 & 0.033 \\
    \multicolumn{1}{l}{mve\_ia} & -179.633 & 6448.617 & -382.103 & c2p\_ia & 11.267 & 296.081 & -0.021 \\
    \multicolumn{1}{l}{pdelay} & 0.142 & 0.980 & 0.069 & cash  & 0.141 & 0.189 & 0.060 \\
    \multicolumn{1}{l}{retvol} & 0.033 & 0.026 & 0.025 & cashpr & -1.853 & 53.433 & -0.873 \\
    \multicolumn{1}{l}{roavol} & 0.023 & 0.044 & 0.011 & convind & 0.136 & 0.343 & 0.000 \\
    \multicolumn{1}{l}{rsup} & 0.018 & 0.202 & 0.021 & d2p   & 0.018 & 0.033 & 0.001 \\
    \multicolumn{1}{l}{baspread} & 0.054 & 0.066 & 0.035 & $\Delta$curr & 0.056 & 0.515 & -0.012 \\
    \multicolumn{1}{l}{std\_dolvol} & 0.864 & 0.395 & 0.837 & $\Delta$quick & 0.078 & 0.611 & -0.010 \\
    \multicolumn{1}{l}{std\_turn} & 3.913 & 7.085 & 1.791 & $\Delta$s2i & 0.120 & 0.901 & 0.017 \\
    \multicolumn{1}{l}{sue} & -0.004 & 0.124 & 0.000 & divi  & 0.030 & 0.170 & 0.000 \\
    \multicolumn{1}{l}{turn} & 1.018 & 1.655 & 0.487 & hire  & 0.084 & 0.332 & 0.019 \\
    \multicolumn{1}{l}{zerotrade} & 1.252 & 3.200 & 0.000 & isc   & 0.082 & 0.090 & 0.052 \\
    \multicolumn{1}{l}{\textbf{\underline{Profitability}}} &       &       &       & lev   & 2.293 & 4.810 & 0.714 \\
    \multicolumn{1}{l}{$\Delta$ato\_ia} & -0.002 & 0.215 & 0.001 & lgr   & 0.234 & 0.748 & 0.086 \\
    \multicolumn{1}{l}{$\Delta$depr} & 0.098 & 0.487 & 0.032 & orgcap & 0.013 & 0.012 & 0.011 \\
    \multicolumn{1}{l}{$\Delta$pm\_ia} & 0.163 & 12.885 & -0.003 & sp    & 2.347 & 3.788 & 1.100 \\
    \multicolumn{1}{l}{$\Delta$tx} & 0.001 & 0.011 & 0.000 & securedind & 0.410 & 0.492 & 0.000 \\
    \multicolumn{1}{l}{depr} & 0.264 & 0.399 & 0.150 & sgr   & 0.183 & 0.558 & 0.106 \\
    \multicolumn{1}{l}{ep} & -0.023 & 0.351 & 0.049 & sin   & 0.009 & 0.092 & 0.000 \\
    \multicolumn{1}{l}{ear} & 0.003 & 0.076 & 0.001 & smi   & -0.040 & 0.716 & 0.021 \\
    \multicolumn{1}{l}{gma} & 0.365 & 0.330 & 0.341 & sma   & -0.038 & 0.570 & 0.004 \\
    \multicolumn{1}{l}{gmms} & -0.067 & 0.856 & -0.002 & sms   & 0.014 & 0.291 & -0.003 \\
    \multicolumn{1}{l}{grltnoa} & 0.080 & 0.131 & 0.063 & tang  & 0.540 & 0.150 & 0.552 \\
    \multicolumn{1}{l}{ms} & 3.731 & 1.599 & 4.000 & \textbf{\underline{Intangibles}} &       &       &  \\
    \multicolumn{1}{l}{nincr} & 0.995 & 1.262 & 1.000 & absacc & 0.089 & 0.095 & 0.068 \\
    \multicolumn{1}{l}{operprof} & 0.806 & 1.099 & 0.687 & acc   & -0.019 & 0.121 & -0.003 \\
    \multicolumn{1}{l}{ps} & 4.183 & 1.653 & 4.000 & age   & 11.712 & 10.119 & 9.000 \\
    \multicolumn{1}{l}{roa} & -0.001 & 0.051 & 0.008 & curr  & 3.163 & 4.680 & 1.982 \\
    \multicolumn{1}{l}{roe} & 0.003 & 0.137 & 0.026 & quick & 2.431 & 4.136 & 1.243 \\
    \multicolumn{1}{l}{roic} & -0.105 & 1.067 & 0.071 & rd    & 0.126 & 0.332 & 0.000 \\
    \multicolumn{1}{l}{tb} & -0.108 & 1.598 & -0.057 & salecash & 52.404 & 159.305 & 10.564 \\
          &       &       &       & saleinv & 22.461 & 60.220 & 7.284 \\
          &       &       &       & salerec & 11.570 & 54.402 & 5.896 \\
    \bottomrule
    \end{tabular*}%

	\scriptsize  Note: This table reports the firm characteristics we use in our empirical study by category. The detail of variable definition is reported in the Appendix. The sample period spans  \sampleperiod
\end{table}

\subsection{Cross section of stock returns}
This subsection mainly illustrates the model performance of the pseudo-SNAP model with the vanilla LSTM (one layer) in terms of out-of-sample prediction and Sharpe ratio, compared with the benchmark models. The robust results for the different number of layers are reported in \autoref{tab:robust_layer}, which shows that the vanilla LSTM has a better out-of-sample performance than the LSTM with multiple hidden layers. The stacked LSTMs, referred to as multiple hidden LSTM layers, have worse performance here because more hidden parameters surpass the dimension of input features\footnote{Following the rule of thumb, we choose the number of neurons between 1/2 to 2/3 of the dimension of input features.}. Therefore, we focus on reporting the results of the pseudo-SNAP model with the vanilla LSTM.

\autoref{tab:R2} reports the predictive $ R^2_{predictive} $ for individual stocks. This metric measures the magnitude of explained variation of excess returns and evaluates the statistical performance of a model. Different types of the pseudo-SNAP model both dominate the Fama-French factor models and the regularized linear model. The nonlinear and flexible functional form enables our model to increase at least 1000\% more in explaining variation of excess returns in the training sample and 200\% more in the validation and test sample. In spite of the fact that pseudo-SNAP3 delivers the highest $ R^2_{predictive} $ of 11.57\% in the training sample, the best model in terms of out-of-sample forecasting is pseudo-SNAP1, which achieves 8.33\% in the validation sample and 4.38\% in the test sample. It is worth noting that the out-of-sample forecasting performance (including the validation and test sample) decays more than 50\% in the existing literature when compared to the performance results in the training sample \citep{chen_deep_2020}. However, our pseudo-SNAP1 out-of-sample forecasting only decays 4.6\% in the validation sample and 49\% in the test sample and it outperforms benchmark models in the out-of-sample performance, which demonstrates that the pseudo-SNAP model with the LSTM network can store long-run information for out-of-sample prediction. We can also see that the masked pseudo-SNAP model suffers from over-fitting issue. In comparing with the masked and unmasked pseudo-SNAP model, the only difference is whether or not to include the alphas in the pseudo-SNAP model, which emphasizes that the economic meaning and correct model speciﬁcation are important elements in the prediction of conditional asset pricing models.
\begin{table}[!htbp]
	\caption{Predictive performance $ R^2_{predictive} $ comparison}

\begin{center}
	
    \begin{tabular}{lccc}
    \toprule
          & \multicolumn{3}{c}{$ R^2_{predictive} $(\%)} \\
\cmidrule{2-4}    Model & Training & Validation & Test \\
    \midrule
    FF3   & -17.76 & -27.30 & -18.28 \\
    FF5   & -21.24 & -33.44 & -22.57 \\
    LASSO & 0.21 & -0.02 & 0.24 \\
    ElasticNet & 0.21 & -0.02 & 0.24 \\
    Ridge & 0.76 & -0.81 & -0.55 \\
    \midrule
    Feedforward Network & -51.61 & -79.16 & -101.70 \\
    \midrule
    pseudo-SNAP\_masked1 & 11.02 & 4.27 & -0.79 \\
    pseudo-SNAP\_masked2 & 10.48 & -12.46 & -2.23 \\
    \midrule
    pseudo-SNAP1(*) & 8.75 & 8.35 & 4.38 \\
    pseudo-SNAP2 & 8.69 & 7.16 & 3.02 \\
    pseudo-SNAP3 & 11.57 & -0.47 & 4.00 \\
    \bottomrule
    \end{tabular}%

\end{center}
\footnotesize Note: This table reports the predictive $ R^2_{predictive} $ for individual level stocks for the Fama-French factors models (FF3, FF5), LASSO, ElasticNet, Ridge, and a variety of pseudo-SNAP masked models on \autoref{eq:model2} and pseudo-SNAP models on \autoref{eq:model1}, respectively.
	\label{tab:R2}%
\end{table}%

\begin{table}[!htbp]
	\caption{Sharpe ratio comparison}
\begin{center}
	
    \begin{tabular}{lccc|ccc}
    \toprule
          & \multicolumn{3}{c}{SR} & \multicolumn{3}{c}{SR\_VW} \\
\cmidrule{2-7}    Model & \multicolumn{1}{l}{Training} & \multicolumn{1}{l}{Validation} & Test  & \multicolumn{1}{l}{Training} & \multicolumn{1}{l}{Validation} & Test \\
    \midrule
    FF3   & -0.03 & -0.43 & 0.01  & -0.08 & -0.43 & -0.04 \\
    FF5   & 0.03  & -0.5  & 0.02  & 0.01  & -0.52 & -0.03 \\
    LASSO & -0.33 & -0.22 & -0.84 & -0.48 & -0.34 & -1.01 \\
    ElasticNet & -0.33 & -0.22 & -0.84 & -0.48 & -0.34 & -1.01 \\
    Ridge & 2.26  & 1.16  & 0.98  & 2.12  & 0.79  & 0.67 \\
    \midrule
    Feedforward Network & 1.63  & 0.49  & 0.80  & 1.46  & 0.40  & 0.67 \\
    \midrule
    pseudo-SNAP\_masked1 &             2.51  &             2.02  &             1.87  &             2.32  &             1.99  &             1.77  \\
    pseudo-SNAP\_masked2 &             2.99  &             2.19  &             1.50  &             3.04  &             2.17  &             1.51  \\
    \midrule
    pseudo-SNAP1(*) &             2.80  &             2.76  &             1.89  &             2.86  &             2.74  &             1.93  \\
    pseudo-SNAP2 &             2.67  &             2.62  &             2.02  &             2.69  &             2.54  &             1.98  \\
    pseudo-SNAP3 &             3.07  &             1.26  &             1.28  &             3.15  &             1.22  &             1.26  \\
    \bottomrule
    \end{tabular}%

\end{center}
	\label{tab:sharpe}%
	\footnotesize Note: This table shows the annualized Sharpe ratio (SR) and its value weighted counterpart (SR\_VW) for long-short portfolios for the Fama-French factor models (FF3, FF5), LASSO, ElasticNet, Ridge, Feedforward network, and a variety of pseudo-SNAP masked models on \autoref{eq:model2} and pseudo-SNAP models on \autoref{eq:model1}, respectively.
\end{table}%

In spite of that the pseudo-SNAP model achieves better performance in prediction compared to other benchmark models, it only describes the statistical performance. Therefore, we evaluate the economic performance of the model through the Sharpe ratio. In \autoref{tab:sharpe} and \autoref{tab:decay}, we find that again the pseudo-SNAP model (masked or unmasked) all outperform the benchmark models with relatively small performance decay. We emphasize the comparison of performance decay between other deep learning approaches and our model in \autoref{tab:decay}. Especially, \cite{chen_deep_2020} introduce Generative Adversarial Networks (GAN) in asset pricing and reports an excellent high monthly Sharpe ratio with 2.68 in the training sample. However, when we look at its performance decay of Sharpe ratio in the out-of-sample, the Sharpe ratio decays 46.64\% and 72.01\% in the validation and test sample, respectively while our pseudo-SNAP1 only decays 1.27\% in the validation sample and pseudo-SNAP2 decays 24.10\% in the test sample. Regardless, overfitting is a common issue in the deep learning neural networks, our model avoids this issue by incorporating the dropout regularization method and LSTM networks to better capture the long-term dependency information of firm characteristics and macroeconomic states. In addition, equal weighted and value weighted Sharpe ratio have close values and, especially, the value-weighted Sharpe ratio of pseudo-SNAP3 stands out with 3.15 while pseudo-SNAP1 achieves better out-of-sample performance. Interestingly, the masked pseudo-SNAP model has a similar magnitude to the Sharpe ratio as the unmasked does, that makes testing for $ \alpha_{it} = 0 $ necessary.

\begin{table}[!htbp]
	\caption{Sharpe ratio performance decay comparison}
	\begin{center}
		
    \begin{tabular}{lccc|ccc}
    \toprule
          & {SR} & \multicolumn{2}{c|}{Performance decay(\%)} & \multicolumn{1}{l}{SR\_VW} & \multicolumn{2}{c}{Performance decay(\%)} \\
\cmidrule{2-7}    Model & \multicolumn{1}{l}{Training} & \multicolumn{1}{l}{Validation} & \multicolumn{1}{c|}{Test} & \multicolumn{1}{l}{Training} & \multicolumn{1}{l}{Validation} & \multicolumn{1}{c}{Test} \\
\midrule
    Feedforward Network & 1.63  & -69.62 & -51.0 & 1.46  & -72.37 & -53.92 \\
    GAN   & 2.68 & -46.64 & -72.01 & \multicolumn{3}{c}{-} \\
    \midrule
    pseudo-SNAP\_masked1 &             2.51  & -19.46& \underline{-25.34 }&                   2.32  & -14.49 & \underline{-23.91} \\
    pseudo-SNAP\_masked2 &             2.99  & -26.77& -49.87 &                   3.04  & -28.69 & -50.43 \\
    \midrule
    pseudo-SNAP1(*) &             2.80  & \underline{-1.27} & -32.52&                   2.86  & \underline{-4.05} & -32.49 \\
    pseudo-SNAP2 &             2.67  & \underline{-1.87} & \underline{-24.10 }&                   2.69  &\underline{ -5.76}& \underline{-26.34} \\
    pseudo-SNAP3 &             3.07  & -58.87 & -58.43 &                   3.15  & -61.39 & -59.91\\
    \bottomrule
    \end{tabular}%

	\end{center}
	\label{tab:decay}%
	\footnotesize Note: This table shows the out-of-sample performance decay of the annualized Sharpe ratio (SR) with equal weight and value weighted (SR\_VW) for long-short portfolios, respectively. The training sample reports the level value while the validation and test sample report the percentage change of the out-of-sample decay rate. Benchmark models include the Fama-French factor models (FF3, FF5), LASSO, ElasticNet, Ridge, Feedforward network, and a variety of pseudo-SNAP masked models on \autoref{eq:model2} and pseudo-SNAP models on \autoref{eq:model1}, and the generative adversarial networks (GAN) model results from \cite{chen_deep_2020}.
\end{table}%

To evaluate the zero mispricing errors, we proceed with the group difference between the residuals of the pseudo-SNAP model and the residuals of the masked counterpart. According to the normality testing results in \autoref{tab:normality}, both the Shaprio-Wilk test and Kolmogrov-Smirnov test reject the null hypothesis of normality, which implies that the traditional $ t $ statistic does not satisfy with its assumption for group difference testing. Hence we use Man-Whitney U test instead and $ t $ test results are also reported as reference in \autoref{tab:alpha_test}. P-value of both Mann-Whitney $ U  $ test and $ t $ are 0 for all of sample division. This testing result implies that the total average of $ \alpha_{it} $ is significantly non-zero as pointed by \cite{kim_arbitrage_2020}.


\begin{table}[!htbp]
	\caption{Mispricing test for null hypothesis: $ \alpha_{it} = 0 $}
\begin{center}

    \begin{tabular}{lcc|cc}
    \toprule
          & \multicolumn{2}{c}{Mann-Whitney U test} & \multicolumn{2}{c}{ t test} \\
\cmidrule{2-5}    Sample & Statistics & P-value & Statistics & P-value \\
    \midrule
    Training & 1661074480147.5 & 0     & -163.15 & 0 \\
    Validation & 23194892686.5 & 0     & -415.81 & 0 \\
    Test  & 56490377983.5 & 0     & -415.45 & 0 \\
    \bottomrule
    \end{tabular}%

\end{center}
	\label{tab:alpha_test}%
	\footnotesize Note: This table reports the testing results for the population mean of $ \alpha_{it} = 0$ through nonparametric test Mann-Whitney $ U $ test and parametric test $ t $ test.
\end{table}%

\subsection{Time series of stock returns}
In this subsection, we construct the arbitrage portfolio as defined earlier. Then we start with running a time-series regression of the arbitrage portfolio's returns on common risk factors. The purpose of this analysis is to check whether the significant alphas are  dependent on these common risk factors. The common risk factors include CAPM, Fama-French 3-factor model, and Fama-French 5-factor model, respectively. In \autoref{tab:ap_reg}, we find that the intercept in the time series regression, alpha, is significant across three asset pricing models while it is only significant at a 10\% level in the FF3 model. Moreover, risk exposure to all factors are significant under the 1\% significance level except cma factor.

\begin{table}[!htbp]
	\caption{Arbitrage portfolio's returns with factors}
	\begin{center}
		
    \begin{tabular}{lccc}
    \toprule
          & CAPM  & FF3   & FF5 \\
\cmidrule{2-4}          & \multicolumn{3}{c}{arbitrage portfolio's return} \\
    \midrule
    alpha & -2.05** & -1.54* & -2.49** \\
          & (0.82) & (0.86) & (1.10) \\
    mktrf & -1.45*** & -1.39*** & -1.11*** \\
          & (0.34) & (0.37) & (0.39) \\
    smb   &       & -1.40*** & -0.99** \\
          &       & (0.45) & (0.45) \\
    hml   &       & -1.20** & -2.00** \\
          &       & (0.57) & (0.79) \\
    rmw   &       &       & 1.58** \\
          &       &       & (0.69) \\
    cma   &       &       & 1.79 \\
          &       &       & (1.39) \\
    \midrule
    $ R^2 $    & 0.09 & 0.14 & 0.17 \\
    Obs.  & 592   & 592   & 592 \\
    \bottomrule
    \end{tabular}%

	\end{center}
	\footnotesize Note: This table reports the OLS regression results of arbitrage portfolio's returns on the risk factors through CAPM, Fama-French 3-factor model, and Fama-French 5-factor model; \stars\stderr
	\label{tab:ap_reg}%
\end{table}%

We further explore the significant properties of mispricing errors. In particular, we dig the arbitrage portfolio deeply through K-Means clustering, which classifies the arbitrage portfolio into 5 clusters\footnote{We use the Elbow method to determine the number of clusters, which are described in the Appendix.}. In particular, we focus on the the time-series properties of the excess returns of the arbitrage portfolio on the out-of-sample (validation and test sample).

\begin{figure}\caption{Out-of-sample arbitrage portfolio 5-Means clustering plot}
   		\label{fig:km_all}
   	\noindent	\begin{subfigure}[b]{0.49\textwidth}
   		\centering
   		\caption{Validation Sample}
   	\label{fig:km_Validation}
   		\includegraphics[width=1\linewidth]{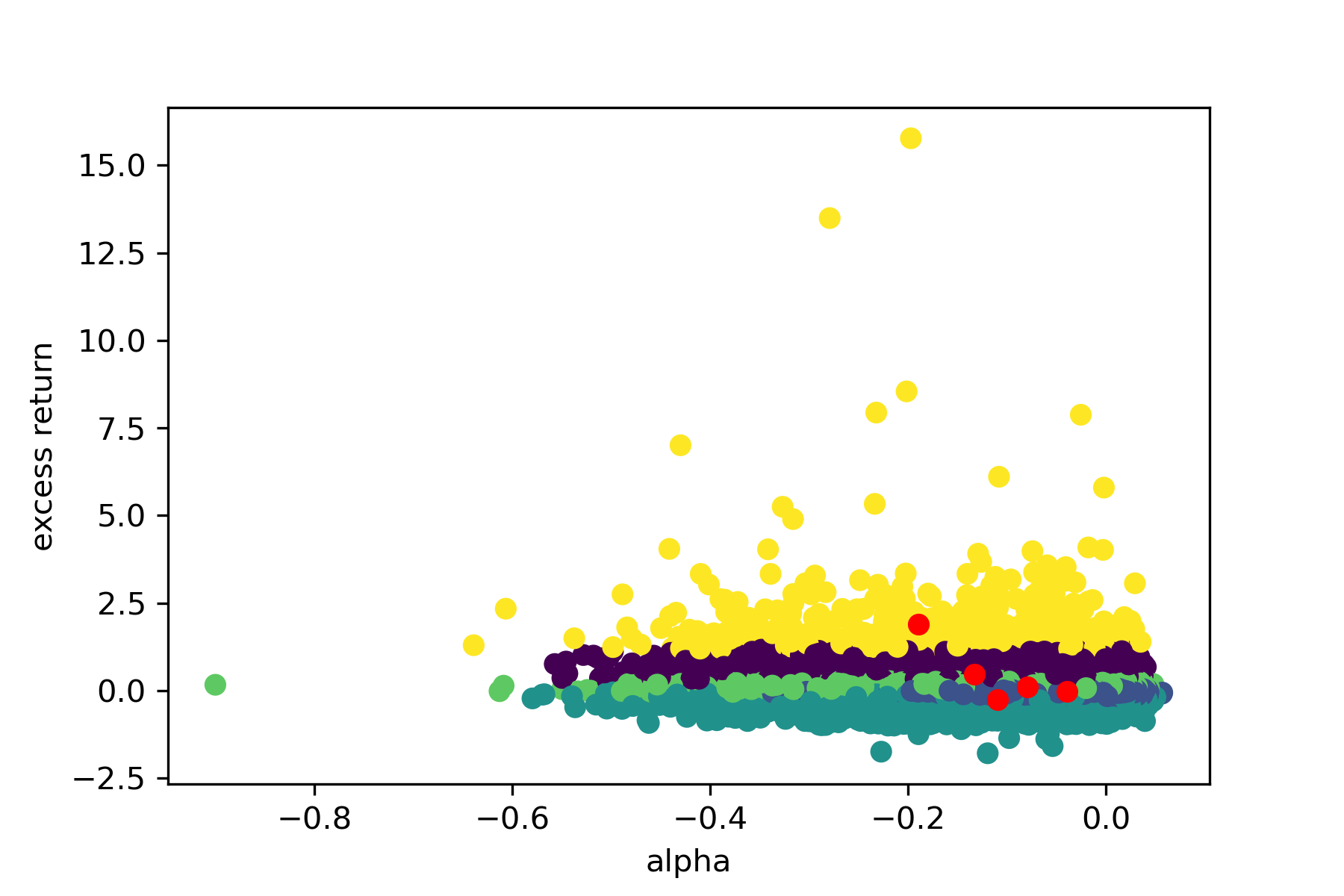}
   		 	\end{subfigure}
   	\noindent	\begin{subfigure}[b]{0.49\textwidth}
   	 	  		\centering
   	 	  		\caption{Test Sample}
   		\label{fig:km_Test}
   			\includegraphics[width=1\linewidth]{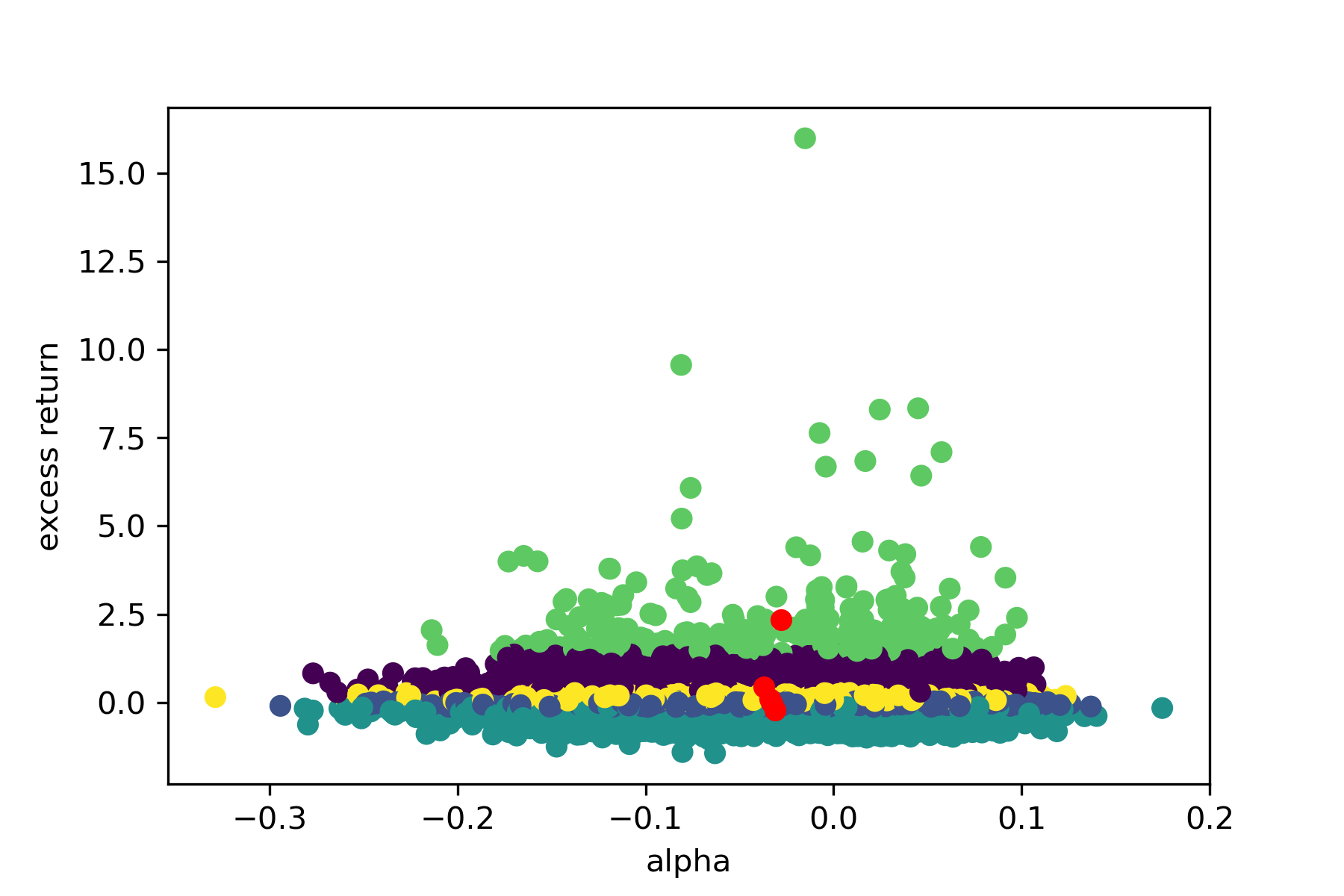}
   	\end{subfigure}
	\footnotesize 
	Note: This figure illustrates the 5-Means clustering for arbitrage portfolios in the validation and test sample. In particular, the red color dots represent the centroid for each cluster.
\end{figure}

A visualization of how arbitrage portfolios are classified through 5-Means clustering is depicted in \autoref{fig:km_all}\footnote{4 extreme excess returns are detected through this figure and we remove them for subsequent time-series discussions.}. From subfigure (a) to (b), the overall groups of $ (\alpha_i, R^e_i) $ are moving from negative values toward positive ones as the centroids are switching to the right direction. In addition, we can see that the clustering of arbitrage portfolios is more dependent on variation of excess returns as the clusterings are distributed vertically toward the dimension of excess returns.

\begin{figure}[H]
   	\caption{Sharpe ratio of arbitrage portfolios clustering}
\begin{center}
\includegraphics[width=0.9\textwidth]{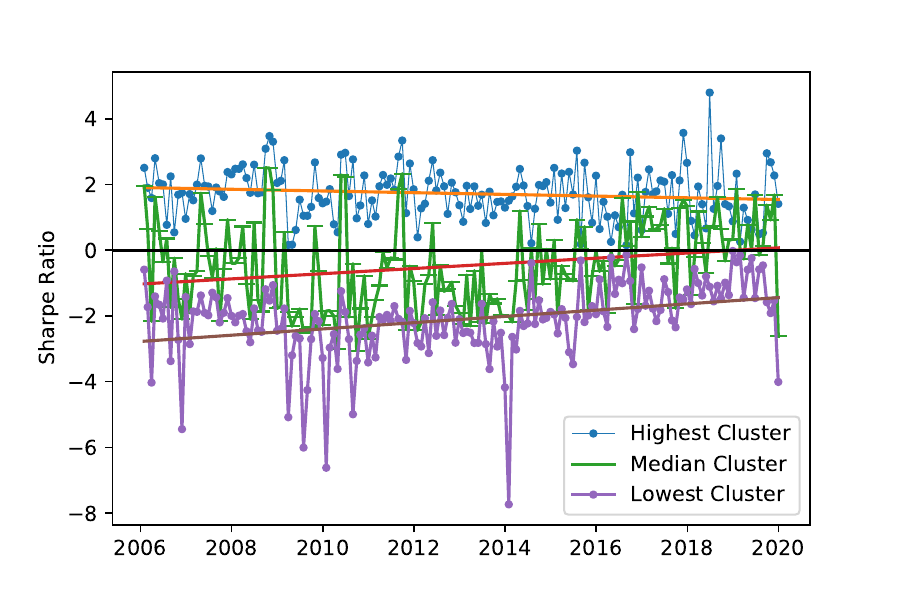}
\end{center}
\label{fig:ts_ap}
\footnotesize Note: This figure shows the monthly Sharpe ratio of the highest, median, and lowest cluster of arbitrage portfolio from January 2006 through December 2019. The regression line is estimated by OLS.
\end{figure}

\autoref{fig:ts_ap} verifies the quote ``In the short run, the market is a voting machine, but in the long run it is a weighing machine" by Benjamin Graham. This quote expresses the distinct stance between short-run and long-run stock market behavior. Transitory news and investors' sentiment are like votes which may drive stock price up or down. In the long run, the fundamentals of a security become clarified over time, and the market begins to weigh its value appropriately. 
As shown in \autoref{fig:ts_ap}, in our out-of-sample, there are always popular and unpopular portfolios with high positive or high negative Sharpe ratio over time,
but the Sharpe ratio difference between the distinct types of portfolios gradually decreases over the long term. Such a long-term phenomenon exactly reveals what matters in the long-run is a firm's actual performance rather than the fickle investing news about its prospects in the short-run. It is worth noting that the Sharpe ratio of the highest cluster of the arbitrage portfolio does not diminish systematically over time and its intercept is significant under the 1\% level and coefficient of $ t $ is insignificant as shown in \autoref{tab:ap_ols}. Similar results are found by \cite{kim_arbitrage_2020} while they do not rule out the negative returns from their arbitrage portfolio and their target variable is the excess return of arbitrage portfolio. Our results re-confirm their findings on out-of-sample Sharpe ratio and point out that such significant average Sharpe ratios are partially un-systematically based on the highest cluster of the arbitrage portfolio and partially systematically based on the median and lowest cluster of the arbitrage portfolio. This finding also has important meanings in practical investment management. Due to the existence of the highest cluster of the arbitrage portfolio, superior stock pickers can achieve higher performance \citep{zambrana_tale_2020}. In addition, portfolio managers and fund managers can buy and hold an investment portfolio based on the highest cluster of the arbitrage portfolio for the long-term investment. This investing strategy is also referred to as the ``Buy-and-Hold Strategy" as Benjamin Graham and Warren Buffett are stalwart fans of this strategy. A potential explanation for both our findings here and this long-term strategy is that there is a strong correlation between the U.S. stock market and U.S. economic growth regardless the economic climate.

\begin{table}[!htbp]
	\caption{Time Series Regression of Sharpe ratio of the Arbitrage Portfolio clustering}
	
	\begin{center}
		
    \begin{tabular}{lccc}
    \toprule
          & (1)   & (2)   & (3) \\
          & high  & median & low \\
    \midrule
    t     & -0.00007 & 0.00022*** & 0.00026*** \\
          & (0.000) & (0.000) & (0.000) \\
    Constant & 3.099*** & -4.650*** & -7.161*** \\
          & (0.808) & (1.334) & (1.095) \\
          \midrule
    Obs     & 168   & 168   & 168 \\
    \bottomrule
    \end{tabular}%

	\end{center}
	\footnotesize Note: This table reports the OLS regression results of out-of-sample Sharpe ratio of the arbitrage portfolio clustering (high, median, and low) on date $ t $; \stars\stderr
	\label{tab:ap_ols}%
\end{table}%

Based on the above findings, our pseudo-SNAP models also provide practical meaning for investors and fund managers. The output of the pseudo-SNAP models are the better out-of-sample predictions, the significant alphas, and the constructed arbitrage portfolios. Although it is hard to predict from alpha how individual stocks react to specific events, investors can tailor a portfolio based on our arbitrage portfolio K-Means clustering. For instance, portfolio managers and fund managers can invest on highest cluster of arbitrage portfolio through Buy-and-Hold Strategy for long-run investment while short selling the lowest cluster in the short-run.

%

\subsection{Characteristic and macroeconomic state importance}

\begin{figure}[H]
	\caption{Characteristic importance rankings}
	\begin{center}
		\includegraphics[width=1\textwidth]{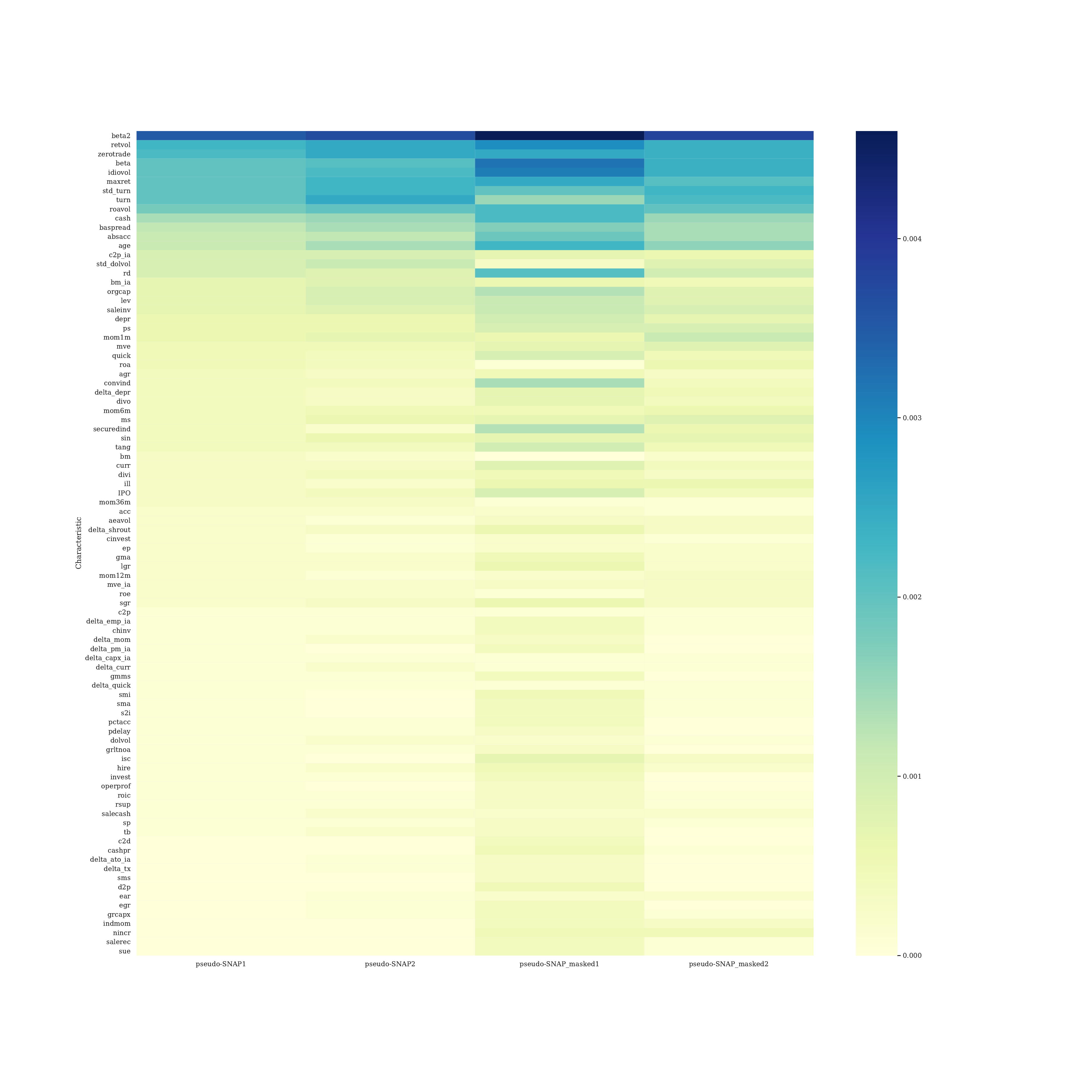}
	\end{center}
	\label{fig:var_importance}
	\footnotesize Note: This heatmap figure shows the importance rankings of all of 90 characteristics over the pseudo-SNAP models and masked pseudo-SNAP models. Each column indicates the most influential firm characteristic (dark blue) to the least influential firm characteristic (light yellow).
	\end{figure}

In this subsection, we investigate which firm characteristic and macroeconomic state carries incremental information for prediction of stock returns. \cite{freyberger_dissecting_2020} utilize the group LASSO method to select characteristics by setting a given characteristic to 0 if it does not contribute to the prediction of returns. However, our pseudo-SNAP model consists of three different neural networks, which are integrated in a pseudo-Siamese network. That makes it hard to identify either firm characteristics or macroeconomic states through a linear regularization method. Estimating the variable importance in the neural network models is a pervasive method to provide explanatory powers from each input feature \citep{olden_accurate_2004}. For simplicity of exhibition, we use firm characteristic as the main example to show how we define and obtain the ranking of characteristics. In particular, we quantify characteristic importance in pseudo-SNAP models through noise-based feature perturbation and its algorithm is summarized as follows:
\begin{enumerate}
\item Compute the predictive values, $ \hat{R}^e_{i,t+1} $, for the original firm characteristics, $ \mathbf{z}_{it} $. For each firm characteristic, $ z_{ikt} $, do steps 2 to 5.
\item Perturb $ z_{ikt} $ by a random normal distribution centered at 0 with scale 0.2 and create $ \mathbf{z}^{*}_{it} $ with only one firm characteristic $ z_{ikt} $ perturbed.
\item Compute the predictions, $ \hat{R}^{*e}_{i,t+1} $ based on the perturbed data $ \mathbf{z}^{*}_{it} $.
\item Compute the Root Mean Square difference between the original $ \hat{R}^e_{i,t+1} $ and the perturbed $ \hat{R}^{*e}_{i,t+1} $.
\item Sort the firm characteristics according to the Root Mean Square difference, and the larger one implies that characteristic is ``more important".
\end{enumerate}
It is worth noting that high dimensional data are highly correlated and often with spurious correlations. This property leads to poor interpretation in high dimensional data modeling, especially in economic or financial modeling. To deal with this poor interpretation issue, we conduct sensitivity analysis to capture the effect of any specific firm characteristic on predicting stock excess returns. The \autoref{fig:var_importance} describes characteristic importance for two pseudo-SNAP models and two alpha-masked ones. Most of characteristic importance hovers around 0 and top 20 characteristics clearly stand out as the influential characteristics. In particular, three categories of firm characteristics serve as the most representative characteristics. The first and biggest group is the trading frictions, which contains 11 influential characteristics: \textit{beta} (market beta), \textit{beta2} (beta squared), \textit{idiovol} (idiosyncratic return volatility), \textit{maxret} (maximum daily return), \textit{retvol} (return volatility), \textit{roavol} (earnings volatility), \textit{baspread} (bid-ask spread), \textit{std\_dolvol} (volatility of liquidity (dollar trading volume)), \textit{std\_turn} (volatility of liquidity (share turnover)), \textit{turn} (share turnover), and \textit{zerotrade} (zero trading days). The second category is intangibles, including 4 firm characteristics: \textit{absacc} (absolute accruals), \textit{age} (\# years since first Compustat coverage), \textit{rd} (R\&D increase), and \textit{saleinv} (Sales to inventory). The last category is value and growth, which involves \textit{bm\_ia} (industry-adjusted book to market), \textit{c2p\_ia} (industry-adjusted cash flow to price), \textit{cash} (cash holdings), \textit{lev} (leverage), and \textit{orgcap} (organizational capital). These 20 firm characteristics are consistent across masked and unmasked pseudo-SNAP models, and interestingly the masked models are affected by relatively more characteristics. This finding demonstrates that the misspecified model without the mispricing errors is unable to accurately explain variation of stock returns with limited firm characteristics but relies on incorporating more firm characteristics to remedy this flaw. But this remedy seems redundant when compared to the pseudo-SNAP model. Therefore, the comparison of characteristic importance rankings between the original pseudo-SNAP model and the alpha-masked model further confirms the importance of considering mispricing errors in conditional asset pricing models.

\begin{figure}[H]
	\caption{Macroeconomic states importance rankings}
	\begin{center}
		\includegraphics[width=1\textwidth]{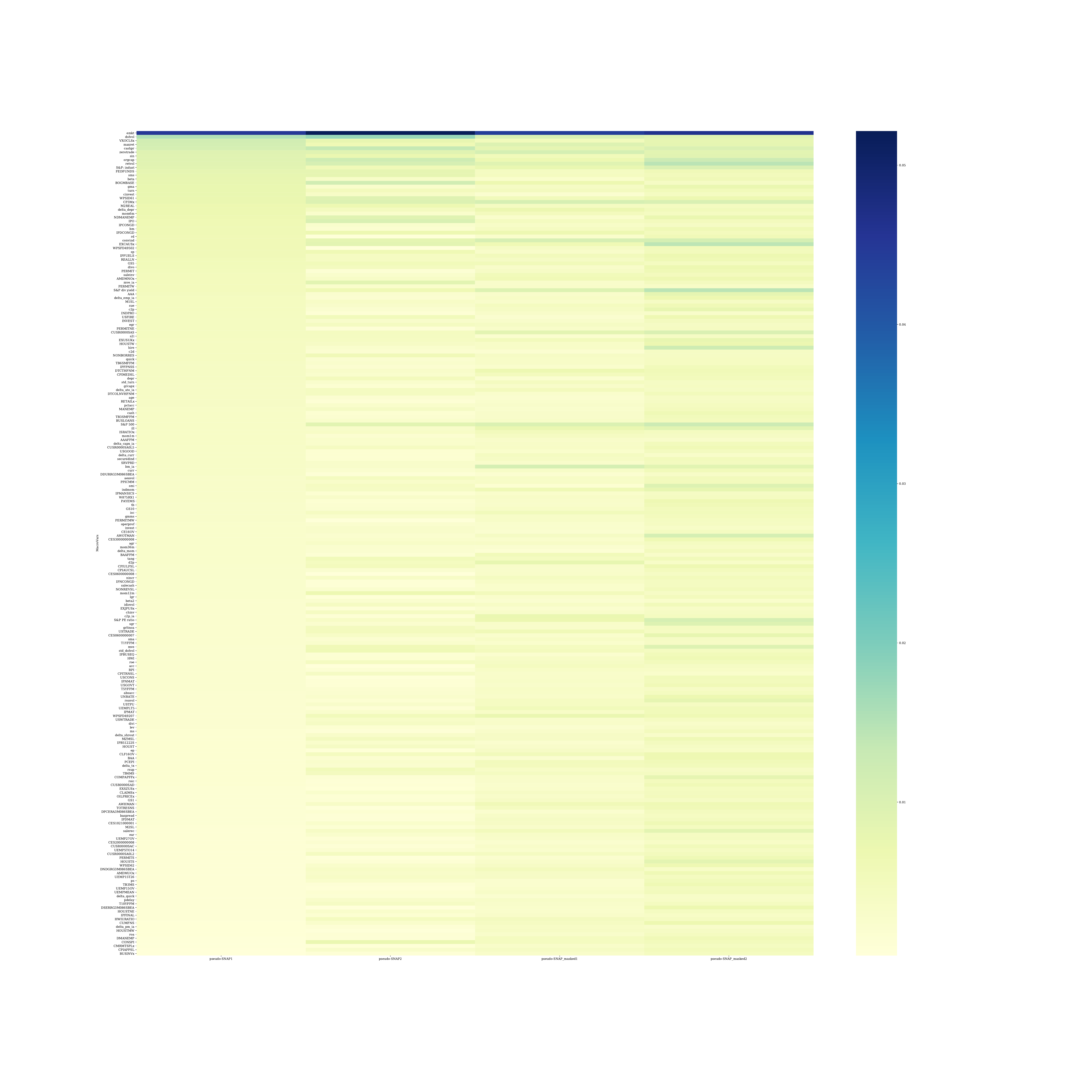}
	\end{center}
	\label{fig:macro_importance}
	\footnotesize Note: This heatmap figure shows the importance rankings of all of 215 macroeconomic states over the pseudo-SNAP models and masked pseudo-SNAP models. Each column indicates the most influential macroeconomic state (dark blue) to the least influential macroeconomic state (light yellow).
\end{figure}

Likewise, \autoref{fig:macro_importance} reports the ranking of important macroeconomic states using the same perturbation method. Different from firm characteristics, the impact of macroeconomic states on predicting stock returns is embedded in risk factor $ \lambda_{t} $ and multiplied by risk exposures, $ \beta_{it} $. Consistent with the existing literature, the market excess factor, \textit{emkt}, stands out as the most important macroeconomic state. Its perturbed root mean square difference is threefold larger than the second most important state, \textit{dovol} (dollar trading volume). In comparing the perturbed root mean square differences between macroeconomic states and firm characteristics, most of macroeconomic states have a larger impact on predicting stock returns than firm characteristics. Especially, 24 of the top 50 important macroeconomic states, are macroeconomic variables while the rest of the average firm characteristics are mostly from the category of trading frictions. This finding suggests that macroeconomic variables, indeed, contribute a salient determinant to predicting stock returns. The top 10 influential macroeconomic variables and their associated groups are shown as follows: Stock market: \textit{VXOCLSx} (The impact of uncertainty shocks), \textit{S\&P: indust} (S\&P's Common Stock Price Index: Industrials); Interest rate:  \textit{FEDFUNDS} (Effective Federal Funds Rate), \textit{CP3Mx} (3-Month AA Financial Commercial Paper Rate); Money and credit: \textit{BOGMBASE} (St. Louis Adjusted Monetary Base), \textit{M2REAL} (Real M2 Money Stock); Prices:  \textit{WPSID61} (PPI: Intermediate Materials): Labor market:  \textit{NDMANEMP} (All Employees: Nondurable goods); Output and income:  \textit{IPCONGD} (IP: Consumer Goods), and \textit{IPDCONGD} (IP: Durable Consumer Goods).

\subsection{Robustness results}
In this subsection, we evaluate pseudo-SNAP models after excluding micro-cap stocks (small stocks) and also compare the performance across different tuning parameters. We find the performance results of the pseudo-SNAP models are robust to the size of firms and tuning parameters. The traditional asset pricing theory exhibits a positive relationship between risk and stock excess returns. This positive risk relationship particularly holds for the large-cap and mid-cap stocks as they are highly liquid and less volatile across markets. However, this relationship may not hold true for small stocks, because low liquidity and high bid-ask spreads of small stocks places limits on trading. Therefore, the stellar Sharpe ratio or significant alphas of micro-cap stocks may not be utilized. It is worth noting that \cite{avramov_machine_2020} find that the machine learning-based return based on the IPCA method will be attenuated after excluding micro-cap stocks. That raises the question whether the pseudo-SNAP model would also suffer from the attenuated performance under the economic restrictions (small stocks excluded).
\begin{table}[!htbp]
	\caption{Performance of pseudo-SNAP models on large market-cap stocks}
	\footnotesize
	\begin{center}
		
    \begin{tabular}{lccc|ccc|ccc}
    \toprule
          & \multicolumn{3}{c}{$ R^2_{predictive} $( \%)} & \multicolumn{3}{c}{SR} & \multicolumn{3}{c}{SR\_VW} \\
\cmidrule{2-10}          & Training & Validation & Test  & Training & Validation & Test  & Training & Validation & Test \\
\midrule
   & \multicolumn{9}{c}{Stock size $ \geq $ 0.001\% of total market cap} \\
    \midrule
pseudo-SNAP 1 & 17.81  & 14.24  & 10.48  &             3.08  &             2.28  &             2.17  &             3.06  &             2.28  &             2.12  \\
pseudo-SNAP 2 & 17.34  & 10.24  & 4.31  &             2.93  &             2.53  &             2.01  &             2.90  &             2.51  &             1.97  \\
pseudo-SNAP 3 & 18.84  & 12.39  & 9.02  &             2.79  &             2.07  &             2.36  &             2.65  &             2.04  &             2.34  \\
\midrule
& \multicolumn{9}{c}{Stock size $ \geq $ 0.002\%  of total market cap} \\
    \midrule
pseudo-SNAP 1 & 19.72 & 13.56 & 12.15 &             3.21  &             2.39  &             2.60  &             3.12  &             2.36  &             2.56  \\
pseudo-SNAP 2 & 20.06 & 16.10 & 12.23 &             3.24  &             2.22  &             2.20  &             3.18  &             2.21  &             2.18  \\
pseudo-SNAP 3 & 19.83 & 10.12 & 8.01 &             3.21  &             1.95  &             2.31  &             3.15  &             1.94  &             2.28  \\
\midrule
& \multicolumn{9}{c}{Stock size $ \geq $ 0.01\%  of total market cap} \\
\midrule
pseudo-SNAP 1 & 23.81  & 23.84  & 17.00  &             2.94  &             3.26  &             2.59  &             2.88  &             3.16  &             2.58  \\
pseudo-SNAP 2 & 23.46  & 23.34  & 16.79  &             3.17  &             3.09  &             2.38  &             3.13  &             3.08  &             2.34  \\
pseudo-SNAP 3 & 23.63  & 22.92  & 16.44  &             3.10  &             2.87  &             2.35  &             3.04  &             2.83  &             2.35  \\
\bottomrule
    \end{tabular}%

	\end{center}
	\footnotesize Note: This table shows the performance of pseudo-SNAP models on large cap stocks in terms of the predictive $ R^2_{predictive} $, the annualized Sharpe ratio (SR) for long-short portfolios and the annualized Sharpe ratio for value-weighted long-short portfolios (SR\_VW). In each subtable, we report the results of 3 models, respectively, with different tuning parameters. There are three cut-offs for small-cap stocks, including 0.001\%, 0.002\%, and 0.01\% of the total market capitalization.
	\label{tab:robust}%
\end{table}%

To evaluate the model performance of pseudo-SNAP model on large-cap and mid-cap stocks, we split the full sample by applying three cut-offs for ruling out small-cap stocks.  0.001\%, 0.002\%, and 0.01\%, represent the stocks with a market-cap larger than 0.001\%, 0.002\%, and 0.01\% of the total market capitalization, respectively. Chopping off the full sample through these three cut-offs separately leaves us 15819 largest stocks, 13154 largest stocks, and 6573 largest stocks.  \autoref{tab:robust} reports the model performance conditional on these three types of largest stocks. In terms of the annual Sharpe ratio, surprisingly our pseudo-SNAP model achieves better performance compared to the one without excluding micro-cap stocks, which is different from the results in the existing literature (performance decreases when excluding micro-cap stocks). The consistent results are also found in terms of the equal weighted Sharpe ratio. As expected, the predictive $ R^2_{predictive} $ are enhanced around three folds compared to the unrestricted sample, even though it is in the out-of-sample. That suggests pseudo-SNAP model is also capable of capturing the total variation in large-cap stocks.

One potential explanation for the stunning and consistent performance of our model conditional on large-cap stocks is that the pseudo-SNAP with long-run memories is able to adapt to the different magnitude of market capitalization and to identify mispriced stocks. In other words, that implies that the stellar performance of our model does not rely on the potential high Sharpe ratios from the micro-cap stocks.

\section{Conclusion} \label{sec:conclusion}
We propose a novel methodology to estimate conditional asset pricing models in a one-step process for individual stocks through utilizing the high dimensional information of financial characteristics and macroeconomic states, while keeping a non-linear and flexible functional form. Our method simultaneously and separately estimates the deep alpha, deep beta, and deep risk factor premia in a pseudo-Siamese network, which is based on the conditional asset pricing models. Our important innovation is that we incorporate the deep learning approaches in the economic model, emphasizing that both of economic meaning and model flexibility are important to out-of-sample performance. This paper also emphasizes the long-term dependency problem and avoids it using the long short-term memory (LSTM).

Our results have a five-fold important contribution to both the literature of asset pricing and practical investment management. First, we propose a pseudo-SNAP model for conditional asset pricing. Our model builds up the bridge between the deep learning literature and conditional asset pricing models, and our pseudo-SNAP models outperform other benchmark models in terms of out-of-sample forecasting and out-of-sample Sharpe ratio. In particular, our model yields a stellar out-of-sample Sharpe ratio above 2 and achieves the smallest performance decay compared to other benchmarks. The performance is robust under economic restrictions. Empirically, we also find that the top 20 influential firm characteristics and macroeconomic variables are important for predicting stock returns. Second, we also discover the importance of economic theory in applying the deep learning approaches. In other words, the correctly specified asset pricing model with appropriate deep learning approaches is the key to achieving state-of-the-art performance. Third, we propose a novel way of testing mispricing errors through simplifying the hypothesis $ \alpha_{it} = 0$ into a group difference testing problem. We provide the testing procedure as well. Fourth, we find that LSTM network not only helps store the long-term dependency to enhance out-of-sample prediction but also keeps the smallest out-of-sample performance decay compared to the benchmark models.
Finally, we implement the arbitrage portfolio K-Means clustering to propose an economic interpretation for Benjamin Graham's quote about the market as a short-term "voting machine" and a long-term "weighing machine."

\clearpage{}
\begin{singlespace}
	\begin{center}
		{\small{}\bibliographystyle{jf}
			\bibliography{DL}
		}{\small\par}
		\par\end{center}
\end{singlespace}

\clearpage{}

\clearpage

\onehalfspacing

\section*{Additional Tables} \label{sec:tab}
\addcontentsline{toc}{section}{Additional Tables}
\begin{table}[!htbp]
	\caption{Normality test for masked pseudo-SNAP residuals and pseudo-SNAP residuals}

\begin{center}
    \begin{tabular}{clcc|cc}
	\toprule
	&       & \multicolumn{2}{c}{Shaprio-Wilk test} & \multicolumn{2}{c}{Kolmogrov-Smirnov test} \\
	\cmidrule{3-6}    Sample & Variable & Statistics & P-value & Statistics & P-value \\
	\midrule
	\multirow{2}[1]{*}{Training} & Masked\_residuals & 0.73  & 0     & 0.36  & 0 \\
	& Residuals & 0.73  & 0     & 0.37  & 0 \\
	\multirow{2}[0]{*}{Validation} & Masked\_residuals & 0.80   & 0     & 0.38  & 0 \\
	& Residuals & 0.74  & 0     & 0.37  & 0 \\
	\multirow{2}[1]{*}{Test} & Masked\_residuals & 0.75  & 0     & 0.38  & 0 \\
	& Residuals & 0.73  & 0     & 0.38  & 0 \\
	\bottomrule
\end{tabular}%
\end{center}

	\label{tab:normality}%
\footnotesize	Note: This table reports the normality test results of the test statistics and P-values using the Shapiro–Wilk test and Kolmogorov–Smirnov test, respectively.
\end{table}%

\begin{table}[!htbp]
	\caption{Robustness check for stacked LSTM(s) with different layers}
	\label{tab:robust_layer}
	\begin{center}

    \begin{tabular}{l|lccc}
    \toprule
    \multicolumn{1}{l}{\# Layers} & Index & Training & Validation & Test \\
    \midrule
    \multirow{2}[2]{*}{L = 1} & $ R^2_{predictive} $    &                                         0.09  &                                                     0.08  &                                                               0.04  \\
          & SR\_VW &                                         2.86  &                                                     2.74  &                                                               1.93  \\
    \midrule
    \multirow{2}[2]{*}{L = 2} & $ R^2_{predictive} $    &                                         0.12  &                                                     0.09  &                                                               0.04  \\
          & SR\_VW &                                         2.35  &                                                     2.26  &                                                               1.63  \\
    \midrule
    \multirow{2}[2]{*}{L = 3} & $ R^2_{predictive} $   &                                         0.12  &                                                     0.07  &                                                               0.03  \\
          & SR\_VW &                                         2.37  &                                                     2.27  &                                                               1.60  \\
    \bottomrule
    \end{tabular}%

	\end{center}
	\footnotesize  Note: This table shows the robust results of $ R^2_{predictive} $ and value-weighted Sharpe ratio (SR\_VW) for our pseudo-SNAP model using different number of layers, $ L $, in \autoref{eq:deep}.
\end{table}

\begin{table}[!htbp]
	\caption{Number of observable data and missing rate for firm characteristics}
	\label{tab:missing_summary}
	\footnotesize

    \begin{tabular*}{1\columnwidth}{lrrlrr}
    \toprule
    Firm Characteristics & Missing Rate, \% & \#Obs & Firm Characteristics & Missing Rate, \% & \#Obs \\
    \midrule
    absacc & 13.71 & 2112884 & ep    & 0     & 2448619 \\
    acc   & 13.71 & 2112884 & gma   & 7.19  & 2272484 \\
    aeavol & 14.25 & 2099747 & grcapx & 16.95 & 2033689 \\
    age   & 0     & 2448619 & grltnoa & 28.5  & 1750824 \\
    agr   & 6.98  & 2277795 & isc   & 0     & 2448609 \\
    baspread & 0     & 2448573 & hire  & 7.2   & 2272334 \\
    beta  & 1.14  & 2420636 & idiovol & 1.14  & 2420636 \\
    beta2 & 1.14  & 2420636 & ill   & 6.24  & 2295891 \\
    bm    & 0     & 2448619 & indmom & 0.25  & 2442495 \\
    bm\_ia & 0     & 2448619 & invest & 10.4  & 2194052 \\
    cash  & 16.5  & 2044601 & IPO   & 0     & 2448619 \\
    c2d   & 3.68  & 2358565 & lev   & 0.27  & 2441905 \\
    cashpr & 1.11  & 2421440 & lgr   & 7.28  & 2270275 \\
    c2p   & 7.75  & 2258804 & maxret & 0     & 2448617 \\
    c2p\_ia & 7.75  & 2258804 & mom12m & 7.97  & 2253512 \\
    $\Delta$ato\_ia & 14.75 & 2087428 & mom1m & 0     & 2448619 \\
    $\Delta$shrout & 7.01  & 2276965 & mom6m & 3.26  & 2368772 \\
    $\Delta$emp\_ia & 7.2   & 2272334 & mom36m & 24.4  & 1851063 \\
    chinv & 9.72  & 2210620 & ms    & 13.04 & 2129422 \\
    $\Delta$mom & 7.97  & 2253512 & mve   & 0     & 2448619 \\
    $\Delta$pm\_ia & 8.42  & 2242549 & mve\_ia & 0     & 2448619 \\
    $\Delta$tx & 18.46 & 1996682 & nincr & 13.04 & 2129422 \\
    cinvest & 18.92 & 1985306 & operprof & 7.2   & 2272228 \\
    convind & 0     & 2448619 & roe   & 13.52 & 2117584 \\
    curr  & 3.46  & 2363957 & roic  & 4.24  & 2344880 \\
    depr  & 4.47  & 2339156 & rsup  & 13.95 & 2106991 \\
    divi  & 6.97  & 2277828 & salecash & 0.89  & 2426719 \\
    orgcap & 26.21 & 1806793 & saleinv & 20.43 & 1948307 \\
    $\Delta$capx\_ia & 9.52  & 2215510 & salerec & 3.6   & 2360367 \\
    $\Delta$curr & 10.54 & 2190619 & securedind & 0     & 2448619 \\
    $\Delta$depr & 11.51 & 2166792 & sgr   & 8.22  & 2247373 \\
    gmms  & 8.23  & 2247146 & sin   & 0     & 2448619 \\
    $\Delta$quick & 11.26 & 2172943 & sp    & 0.26  & 2442232 \\
    smi   & 25.84 & 1815897 & std\_dolvol & 6.37  & 2292566 \\
    sma   & 10.85 & 2182995 & quick & 4.12  & 2347767 \\
    sms   & 22.28 & 1903020 & rd    & 6.97  & 2277828 \\
    s2i   & 26.71 & 1794692 & retvol & 0     & 2448542 \\
    pctacc & 13.71 & 2112872 & roa   & 16.05 & 2055651 \\
    pdelay & 1.14  & 2420607 & roavol & 27.19 & 1782723 \\
    ps    & 6.97  & 2277828 & std\_turn & 6.16  & 2297703 \\
    divo  & 6.97  & 2277828 & sue   & 13.82 & 2110099 \\
    dolvol & 6.77  & 2282914 & tang  & 4.34  & 2342379 \\
    d2p   & 0.22  & 2443207 & tb    & 11.74 & 2161092 \\
    ear   & 13.11 & 2127565 & turn  & 6.72  & 2284050 \\
    egr   & 6.99  & 2277539 & zerotrade & 6.24  & 2295921 \\
    \bottomrule
    \end{tabular*}%

\scriptsize  Note: This table summarizes the observable number of firm characteristics-month data over \sampleperiod with the missing rate. This sample contains all common stocks listed on three major stock exchanges: NYSE, AMEX, and NASDAQ, with available annual and quarterly accounting data from Compustat and individual stock returns from the Center for Research in Securities Prices (CRSP).
\end{table}


\clearpage

\onehalfspacing

\section*{Characteristic Description} \label{sec:chars_des}
\begin{table}[!htbp]
	\caption{Characteristic definition}
	\scriptsize
	
    \begin{tabular}{lll}
    \toprule
    \textbf{Acronym} & \textbf{Definition of the characteristic} & \textbf{Reference} \\
    absacc & abs(\textit{acc}) & Bandyopadhyay, Huang, and Wirjanto (2010) \\
    acc   & (\textit{ib }- \textit{oancf})/\textit{at}, if \textit{oancf} is nonmissing;  & Sloan (1996) \\
          & else set to  &  \\
          & $\Delta$\textit{act} - $\Delta$\textit{che} - $\Delta$\textit{lct} + $\Delta$\textit{dlc} + $\Delta$\textit{txp}-\textit{dp} &  \\
          & \textit{ib}: Annual income before extraordinary items &  \\
          & \textit{oancf}: operating cash ﬂows, bar\_at: average total assets &  \\
    aeavol & (\textit{vol} - $ \bar{vol} $)/\textit{rdq} & Lerman, Livnat,  \\
          & \textit{vol}: average daily trading volume  & and Mendenhall (2008) \\
          & \qquad for 3 days around earnings announcement &  \\
          & $ \bar{vol} $: average daily volume for 1-month ending  &  \\
          &                \qquad 2 weeks before earnings announcement &  \\
          & \textit{rdq}: 1-month average daily volume &  \\
          & \textit{Earnings announcement day from Compustat quarterly} &  \\
    age   & Number of years since ﬁrst Compustat coverage & Jiang, Lee, and Zhang (2005) \\
    agr   & Annual $\Delta$\textit{at}& Cooper, Gulen, and Schill (2008) \\
    baspread & Monthly average of daily bid-ask spread/average of daily spread & Amihud and Mendelson (1989) \\
    beta  & Estimated market beta & Fama and MacBeth (1973) \\
    beta2 & beta squared & Fama and MacBeth (1973) \\
    bm    & Book value of equity / market capitalization & Rosenberg, Reid, and Lanstein (1985) \\
    bm\_ia & Industry adjusted book-to-market ratio & Asness, Porter, and Stevens (2000) \\
    cash  & \textit{che} / $ \bar{at} $ &  Palazzo (2012) \\
    c2d   & \textit{(ib+dp)/lt} & Ou and Penman (1989) \\
          & \textit{ib}: Earnings before depreciation, dp: extraordinary items &  \\
          & \textit{lt}: average total liabilities ($ \bar{lt} $) &  \\
    cashpr & (market cap + long-term debt - total assets) / \textit{che} & Chandrashekar and Rao (2009) \\
          & \textit{che}: cash and equivalents &  \\
    c2p   & Operating cash flows / market cap & Desai, Rajgopal, and Venkatachalam (2004) \\
    c2p\_ia & Industry adjusted \textit{c2p} & Asness, Porter and Stevens (2000) \\
    $\Delta$ato\_ia & $\Delta$sale / average total assets ($ \bar{at} $) & Soliman (2008) \\
    $\Delta$shrout & $\Delta$shares outstanding (\textit{csho}) & Pontiff and Woodgate (2008) \\
    $\Delta$emp\_ia & Industry-adjusted $\Delta$number of employees & Asness, Porter, and Stevens (1994) \\
    chinv & $\Delta$inventory / $ \bar{at} $ & Thomas and Zhang (2002) \\
    $\Delta$mom & $ R_{t-6}^{t-1} - R_{t-12}^{t-7} $ & Gettleman and Marks (2006) \\
    $\Delta$pm\_ia & $\Delta$\textit{ib} / \textit{sale} & Soliman (2008) \\
    $\Delta$tx & $\Delta$total taxes (\textit{txtq}) from quarter \textit{t-4} to \textit{t} & Thomas and Zhang (2011) \\
    cinvest & $\Delta$PP\&E\_{quarter} / saleq - $ \bar{PP\&E} $; if saleq = 0, scale by 0.01 & Titman, Wei, and Xie (2004) \\
    convind & $ \mathbbm{1}_{\text{convertible debt obligations}} $& Valta (2016) \\
    curr  & assets / liabilities & Ou and Penman (1989) \\
    depr  & Depreciation / PP\&E & Holthausen and Larcker (1992) \\
    divi  &$ \mathbbm{1}_{\text{company pays dividends but not in prior year}}$ & Michaely, Thaler, and Womack (1995) \\
    orgcap & Capitalized SG\&A expenses & Eisfeldt and Papanikolaou (2013) \\
    $\Delta$\textit{capx\_ia} & $\Delta$capital expenditures (\textit{capx}) & Abarbanell and Bushee (1998) \\
    $\Delta$curr & $\Delta$\textit{curr} & Ou and Penman (1989) \\
    $\Delta$depr & $\Delta$\textit{depr} & Holthausen and Larcker (1992) \\
    gmms  & $\Delta$gross margin (\textit{sale-cogs}) - $\Delta$\textit{sale} & Abarbanell and Bushee (1998) \\
    $\Delta$quick & $\Delta$\textit{quick }& Ou and Penman (1989) \\
    smi   & $\Delta$\textit{sale} - $\Delta$\textit{invt} & Abarbanell and Bushee (1998) \\
    sma   & $\Delta$\textit{sale }- $\Delta$\textit{rect }& Abarbanell and Bushee (1998) \\
    sms   & $\Delta$\textit{sale} - $\Delta$SG\&A & Abarbanell and Bushee (1998) \\
    s2i   & $\Delta$\textit{saleinv} & Ou and Penman (1989) \\
    pctacc & (\textit{ib }- \textit{oancf} )/abs(\textit{ib}), if oancf is nonmissing;  & Hafzalla, Lundholm,  \\
          & else set to  & and Van Winkle (2011) \\
          & $\Delta$\textit{act} $\Delta$\textit{che} - $\Delta$\textit{lct} + $\Delta$\textit{dlc} + $\Delta$\textit{txp-dp} &  \\
          & if \textit{ib} = 0, then set \textit{ib}= 0.01 &  \\
    pdelay & The ratio of variation in weekly returns for 36 months  & Hou \& Moskowitz (2005) \\
          & ending in month t  explained by 4 lags of weekly market returns  &  \\
          & incremetnal to contemporaneous market return &  \\
    ps    & $ \sum^{9} \mathbbm{1}_{health_i}$ & Piotroski (2000) \\
    \bottomrule
    \end{tabular}%

	\label{tab:chars_description1}%
	\footnotesize	(\textit{continued})
\end{table}%

\begin{table}[!htbp]
	\scriptsize
	
    \begin{tabular}{lll}
    \toprule
    \textbf{Acronym} & \textbf{Definition of the characteristic} & \textbf{Reference} \\
    divo  & $ \mathbbm{1}_{\text{company does not pay dividend but did in prior year}} $ & Michaely, Thaler, and Womack (1995) \\
    dolvol &$  log(\textit{volume})\times price_{t-2}^2  $& Chordia, Subrahmanyam, and Anshuman (2001) \\
    d2p   & \textit{dvt} / market cap & Litzenberger and Ramaswamy (1982) \\
    ear   & Sum of daily returns in three days around earnings announcement & Kishore et al. (2008) \\
    egr   & $\Delta$ book value of equity (\textit{ceq}) & Richardson et al. (2005) \\
    ep    & \textit{ib / market cap} & Basu (1977) \\
    gma   & (\textit{revt} - \textit{cogs}) / L.\textit{at}  & Novy-Marx (2013) \\
    grcapx & $\Delta$$ capx_{t-2}^{t} $ & Anderson and Garcia-Feijoo (2006) \\
    grltnoa & Growth in long-term net operating assets & Fairﬁeld, Whisenant, and Yohn (2003) \\
    isc   & sum of squared percent of sales in industry for each company & Hou and Robinson (2006) \\
    hire  & $\Delta$ number of \textit{emp }& Bazdresch, Belo, and Lin (2014) \\
    idiovol & Standard deviation of residuals of weekly returns on   & Ali, Hwang, and Trombley (2003) \\
          & \qquad weekly equal weighted mkt returns for 3 years &  \\
    ill   & abs(\textit{return}) / \textit{dollar volume} & Amihud (2002) \\
    indmom & Equal weighted average industry 12-month returns & Moskowitz and Grinblatt (1999) \\
    invest & ($\Delta$ gross property, plant, and equipment (\textit{ppegt}) + $\Delta$\textit{invt}) / L.\textit{at} & Chen and Zhang (2010) \\
    IPO   & $ \mathbbm{1}_{\text{first year available}} $ & Loughran and Ritter (1995) \\
    lev   & \textit{lt / market cap} & Bhandari (1988) \\
    lgr   & $\Delta$ total liabilities (\textit{lt}) & Richardson et al. (2005) \\
    maxret & Maximum daily return from returns during calendar month (\textit{t-1}) & Bali, Cakici, and Whitelaw (2011) \\
    mom12m & 11-month cumulative returns ending one month before month end & Jegadeesh (1990) \\
    mom1m & 1-month cumulative return & Jegadeesh and Titman (1993) \\
    mom6m & 5-month cumulative returns ending \textit{t-1 }& Jegadeesh and Titman (1993) \\
    mom36m & Cumulative returns from months \textit{t-36} to \textit{t-13 }& Jegadeesh and Titman (1993) \\
    ms    & $ \sum^8 \mathbbm{1}_{\text{fundamental performance }} $& Mohanram (2005) \\
    mve   &$  log(\textit{market cap}) $ & Banz (1981) \\
    mve\_ia & Industry-adjusted market capitalization & Asness, Porter, and Stevens (2000) \\
    nincr & \# consecutive quarters ($ <= $8 quarters) with an increase in earnings & Barth, Elliott, and Finn (1999) \\
    operprof & (revenue - cost of goods sold - SG\&A exp. - interest exp.) / L.equity & Fama and French (2015) \\
    roe   & Earnings before extraordinary items / L.equity & Hou, Xue, and Zhang (2015) \\
    roic  & (\textit{ebit }- \textit{nopi}) / (\textit{ceq }+ \textit{lt }- \textit{che}) & Brown and Rowe (2007) \\
    rsup  & (sales from quarter t - sales from quarter t-4) / market cap & Kama (2009) \\
    salecash & \textit{sale / che} & Ou and Penman (1989) \\
    saleinv & \textit{sale / inv} & Ou and Penman (1989) \\
    salerec & \textit{sale} / accounts receivable & Ou and Penman (1989) \\
    securedind & $ \mathbbm{1}_{\text{company has secured debt obligations}} $ & Valta (2016) \\
    sgr   & $\Delta$\textit{sale} & Lakonishok, Shleifer, and Vishny (1994) \\
    sin   & $ \mathbbm{1}_{\text{classification is in smoke or tobacco, beer or alcohol or gaming}} $& Hong \& Kacperczyk (2009) \\
    sp    & \textit{sale / market cap} & Barbee, Mukherji, and Raines (1996) \\
    std\_dolvol & Monthly standard deviation of daily dollar trading volume & Chordia, Subrahmanyam, and Anshuman (2001) \\
    quick & (current assets - invt) / current liabilities & Ou and Penman (1989) \\
    rd    & $ \mathbbm{1}_{\text{R\&D expense over total assets has an increase greater than 5\%}}  $& Eberhart, Maxwell, and Sddique (2004) \\
    retvol & Standard deviation of daily returns from month \textit{t-1 }& Ang et al. (2006) \\
    roa   &\textit{ ibq} / L.\textit{atq}, in a quarter & Balakrishnan, Bartov, and Faurel (2010) \\
    roavol & std(\textit{ibq}) / L.$ \bar{atq} $,  & Francis et al. (2004) \\
    std\_turn & Monthly standard deviation of daily share turnover & Chordia, Subrahmanyam, and Anshuman (2001) \\
    sue   & Unexpected quarterly earnings / maket cap & Rendelman, Jones, and Latane (1982) \\
    tang  & \textit{cash}+0.715*receivables+0.547*\textit{invt}+0.535*PPE/\textit{at} & Almeida and Campello (2007) \\
    tb    & Tax income / maximum federal tax rate / \textit{ib} & Lev and Nissim (2004) \\
    turn  & Average monthly trading volume  & Datar, Naik, and Radcliffe (1998) \\
          & for most recent 3 months  &  \\
          & scaled by number of shares outstanding in current month &  \\
    zerotrade & Turnover weighted number of 0 trading days for recent 1 month & Liu (2006) \\
    \bottomrule
    \end{tabular}%

	\label{tab:chars_description2}%
	\footnotesize Note: The definitions and descriptions of our firm characteristics are based on the lists of firm characteristics collected by \cite{green_characteristics_2017}.
\end{table}%

\clearpage
\section*{Figures} \label{sec:fig}
\addcontentsline{toc}{section}{Figures}

   \begin{figure}[!htbp]
   	\caption{Elbow method: Detection of the number of clusters}

	\noindent\begin{subfigure}[b]{1\textwidth} 
\begin{center}
\includegraphics[width=0.5\linewidth]{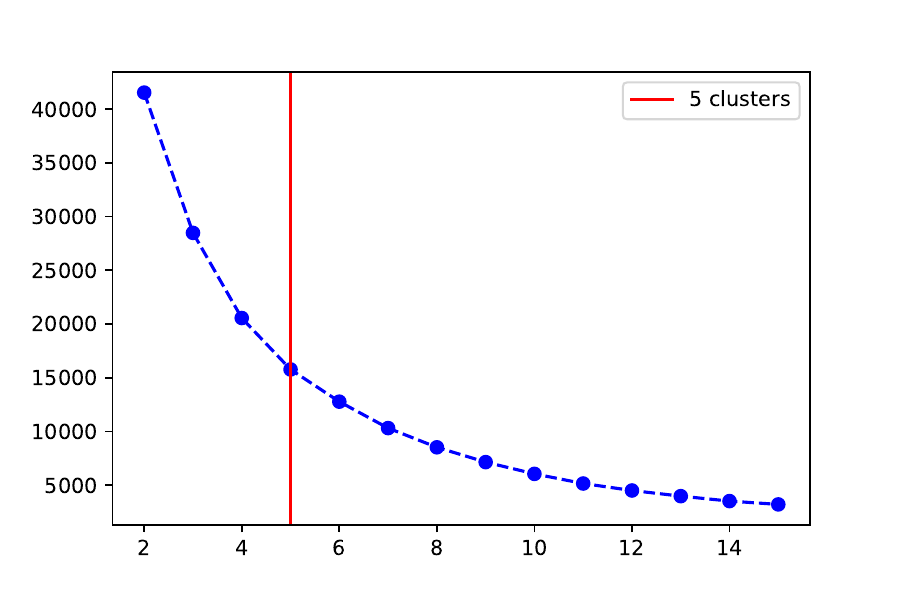}
\caption{Training sample}
\label{fig:detection_train}
\end{center}
	\end{subfigure}\hfil
\noindent\begin{subfigure}[b]{1\textwidth}
\begin{center}
\includegraphics[width=0.5\linewidth]{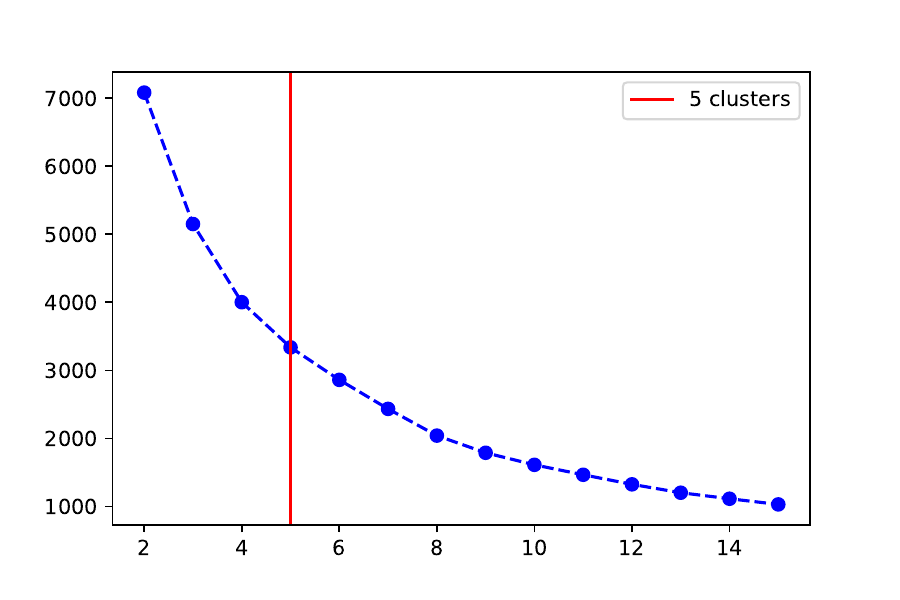}
\caption{Validation sample}
\label{fig:detection_validate}
\end{center}
	\end{subfigure}\hfil
	\noindent\begin{subfigure}[b]{1\textwidth}
	\begin{center}
	\includegraphics[width=0.5\linewidth]{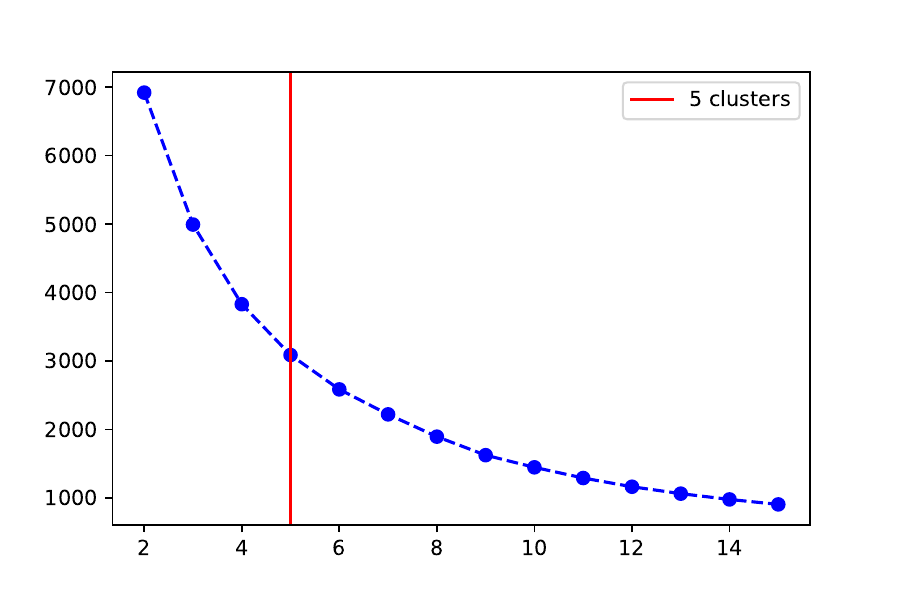}
\caption{Test sample}
\label{fig:detection_test}
	\end{center}
	\end{subfigure}\hfil
\\
		\footnotesize
	Note: This figure depicts the sum of squared error for the range of clusters from 2 to 15. The ``Elbow Method" is used here to choose the number of clusters for K-Means algorithm.
	As indicated, the error rate starts to slow down decreasing at 5.
\end{figure}



\clearpage
\section*{Benchmark Models} \label{sec:benchmark}
\addcontentsline{toc}{section}{Benchmark Models}

\begin{enumerate}
\item  \textbf{Fama French Factor Models}: We refer Fama French three-factor model and Fama French five-factor model separately as ``FF3" and ``FF5". The three-factor model is the market excess return, SMB, and HML, and the five-factor model is the market excess return, SMB, HML, RMW, and CMA (Please visit Fama French library for more details of factors description).

\item \textbf{Linear Regularized Models}: preference for different  eligible capacities.
\begin{equation*}
	\hat{\beta} = \arg \min_{\beta} \sum_{i} L(y_i - f(x_i, \beta)) + \lambda P(\beta)
\end{equation*}
For loss function $ L(.) $ and penalty function $ P(.) $,
\begin{itemize}
	\item  $ P(\beta) = ||\beta||_{1}$: \textbf{Lasso}
	\item  $ P(\beta) = ||\beta||_{2}^2$: \textbf{Ridge regression}
	\item  $ P(\beta) = (1-\alpha)||\beta||_{1}    +\alpha ||\beta||_{2}^2$: \textbf{Elastic net}
\end{itemize}
\item  \textbf{Feedforward Neural Network (FFN)} or \textbf{Multilayers Perceptron (MLP)}

\begin{itemize}
\item[] \begin{neuralnetwork}[height=4]
	\newcommand{\x}[2]{$x_#2$}
	\newcommand{\y}[2]{$\hat{y}_#2$}
	\newcommand{\hfirst}[2]{\small $h^{(1)}_#2$}
	\newcommand{\hsecond}[2]{\small $h^{(2)}_#2$}
	\inputlayer[count=3, bias=true, title=Input\\layer, text=\x]
	\hiddenlayer[count=4, bias=false, title=Hidden\\layer 1, text=\hfirst] \linklayers
	\outputlayer[count=1, title=Output\\layer, text=\y] \linklayers
\end{neuralnetwork}
\item[] Note: This figure briefly summarizes the shallow feedforward network for conditional asset pricing models. The input layer represents high dimensional firm characteristics at current time $ t $, and the output layer denotes the one-month forward stock return at $ t+1 $. The hidden layer bridges between the inputs and the output through the nonlinear activation function.
\end{itemize}
\end{enumerate}

\clearpage

\end{document}